\documentclass[%
reprint,
superscriptaddress,
 amsmath,amssymb,
 aps,
prb,
floatfix,
]{revtex4-1}

\usepackage{graphicx}% Include figure files
\usepackage{dcolumn}% Align table columns on decimal point
\usepackage{bm}% bold math
\usepackage[bookmarks=true,colorlinks=true,linkcolor={blue},citecolor=blue,urlcolor=blue]{hyperref} %This line was 
\usepackage[mathlines]{lineno}% Enable numbering of text and display math
\usepackage{longtable}	% added by suleiman
\usepackage{tabularx}	% added by suleiman
\usepackage{array}		% added by suleiman
\usepackage{multirow}	% % added by suleiman
\usepackage{array}		% To change line spacing within tables
\usepackage{longtable}
\usepackage{subfigure}
\begin{document}

% number in the USPPer righthand corner of the title page in preprint mode.
%\preprint{}

\title{Structural, electronic and optical characterization of bulk platinum nitrides:\\a first-principles study}


\author{Mohammed S. H. Suleiman}
\email[Corresponding author: ]{suleiman@aims.ac.za}
%\thanks{}
%\altaffiliation[Also at: ]{Department of Physics, Sudan University of Science and Technology, Khartoum, Sudan}
\affiliation{School of Physics, University of the Witwatersrand, Johannesburg, South Africa.}
\affiliation{Department of Physics, Sudan University of Science and Technology, Khartoum, Sudan.}

\author{Daniel P. Joubert}
\homepage[Homepage: ]{http://www.wits.ac.za/staff/daniel.joubert2.htm}
\affiliation{School of Physics, University of the Witwatersrand, Johannesburg, South Africa.}

\date{\today}

\begin{abstract}
We present a detailed quantum mechanical non empirical DFT investigation of the energy-optimized geometries, phase stabilities and electronic properties of bulk Pt$_3$N, PtN and PtN$_2$ in a set of twenty different crystal structures. Structural preferences for these three stoichiometries were analyzed and equilibrium structural parameters were determined. We carefully investigated the band-structure and density of states of the relatively most stable phases. Further, GW$_{0}$ calculations within the random-phase approximation (RPA) to the dielectric tensor were carried out to derive their frequency-dependent optical constants of the most likely candidates for the true crystal structure. Obtained results were comprehensively compared to previous calculations and to experimental data.
\end{abstract}
%\keywords{gold nitrides, crystal structure, electronic properties, band structure, density of states, optical properties, DFT, GW approximation.}
\maketitle
\tableofcontents	% This line was added by M Suleiman
% =============================================================================================================
\section{\label{Introduction}Introduction}
% =============================================================================================================
%
% -------------------------------------------------------------------------------------------------------------
%\subsection{\label{Experimental Situation}Experimental Situation}
% -------------------------------------------------------------------------------------------------------------
% Considering chemical stoichiometry:
Platinum is known to form simple binary compounds with other elements (e.g. PtF$_4$, PtI$_2$, PtO and PtS) \cite{first_PtN_2004_exp}. However, platinum had not been known to form crystalline solid nitride, but other forms of platinum nitrides (e.g. PtN \cite{first_PtN_2004_exp,doi:10.1021/jp011542l}, PtN$_2$ \cite{first_PtN_2004_exp}, (PtN)$_2$ \cite{first_PtN_2004_exp,doi:10.1021/jp011542l} and Pt$_2$N \cite{doi:10.1021/jp011542l}) had been observed.

In January 2004, Soto \cite{G_Soto_PtN_x_2004_exp} reported the preparation of platinum thin films containing up to $\sim 14 \; \text{at.} \; \%$ nitrogen. The study concluded that platinum can form an incipient nitride phase with composition near to Pt$_6$N. Few months later, in May 2004, Gregoryanz and co-workers \cite{first_PtN_2004_exp} published the discovery and characterization of solid crystalline platinum mono-nitride for the first time. The synthesis was carried out above $45 \; \text{GPa}$ and $2000 \; K$ but with complete recovery of the product at room pressure and temperature. The produced samples have a very high bulk modulus leading to important implications in high-pressure physics and technology. The 1:1 stoichiometry was assigned to the new nitride, and according to their XRD measurements, Gregoryanz et al. proposed three structures: B1, B3 and B17 (for description of the structures see sub-section \ref{Stoichiometries and Crystal Structures} below), all based on the Pt fcc sub-lattice. Due to some considerations, B1 and B17 were ruled out and B3 was assigned to the new product.

In addition to the well-crystallized and highly ordered regions, a common feature in the synthesized platinum nitrides is the presence of sub- or/and super-stoichiometric phases containing N or Pt vacancies and residual non-stoichiometric material distributed throughout the samples \cite{first_PtN_2004_exp,PtN_n_PtN2_2006_exp_n_comp}.
%
% -------------------------------------------------------------------------------------------------------------
%\subsection{\label{Theoretical Investigations}Theoretical Investigations}
% -------------------------------------------------------------------------------------------------------------

The work of Gregoryanz et al. \cite{first_PtN_2004_exp} has stimulated many further theoretical studies \cite{PtN_n_PtN2_2005_July_comp_5+,positive_Ef_n_PtN2_2006_comp_Scandolo,PtN_PtN2_2009_comp} as expected by Gregoryanz and co-workers \cite{first_PtN_2004_exp} themselves. However, theoretical work showed that PtN(B3) is elastically and thermodynamically unstable (see sub-section \ref{EOS and Relative Stabilities}  below). Accordingly, claiming that large errors are generally inevitable in the used experimental characterization methods \cite{PtN_PtN2_2005_comp,positive_Ef_n_PtN2_2006_comp_Scandolo}, and due to other paradoxical facts \cite{Mysterious_Platinum_Nitride_2006_comp} in the original experiment by Gregoryanz et al. \cite{first_PtN_2004_exp}, theoreticians questioned the chemical stoichiometry and the crystal structure of this new material and started to investigate other possibilities \cite{Mysterious_Platinum_Nitride_2006_comp,PtN_PtN2_2005_comp,PtN_2006_comp}. Moreover, the experimentally reported \cite{first_PtN_2004_exp} high bulk modulus of the platinum nitride has not been reproduced by any reliable calculations and its mechanism is still an unclear open problem \cite{PtN_n_PtN2_2005_July_comp_5+,Mysterious_Platinum_Nitride_2006_comp,PtN_2006_comp}.

These investigations led to a kind of consensus that the compound does not crystallize in the proposed PtN(B3) phase \cite{PtN_n_PtN2_2006_exp_n_comp}, but the true stoichiometry and the true crystal structure have become now a matter of debate \cite{Mysterious_Platinum_Nitride_2006_comp,positive_Ef_n_PtN2_2006_comp_Scandolo}.

In an apparent attempt to respond to this debate, Crowhurst et al. \cite{PtN_n_PtN2_2006_exp_n_comp} managed, in 2006, to to reproduce and characterize platinum nitride. Combining theory with their own observed Raman spectrum, they came up with a conclusion to propose PtN$_2$(C2) and rejected PtN(B3), proposed by the first platinum nitride synthesizers \cite{first_PtN_2004_exp}, and PtN$_2$(C1), proposed in some theoretical works. Like the first proposed structures \cite{first_PtN_2004_exp}, C1 \cite{PtN_PtN2_2005_comp} and C2 \cite{PtN_n_PtN2_2006_exp_n_comp} structures have the fcc sub-lattice of the metal.

Despite the considerable number of the subsequent theoretical studies, the discrepancy between theory and experiment in the structural and the physical properties of this nitride is not yet satisfactorily understood. Nevertheless, many transition metals can form more than one nitride \cite{StructuralInChem}. Thus, it is of interest to know if platinum can form nitrides with different stoichiometries and/or crystal structures other than those proposed by the first platinum nitride synthesizers and other researchers.
%
% -------------------------------------------------------------------------------------------------------------
%\subsection{\label{Present Work}Present Work}
% -------------------------------------------------------------------------------------------------------------

In the present work, we present a comprehensive first-principles calculations of the equation of state, possible pressure-induced phase transitions, electronic and optical properties of crystalline Pt$_3$N, PtN and PtN$_2$ in a total of twenty different -previously proposed and new hypothetical- structural modifications. The work partly aims to solve some of the reported discrepancies. In addition, to the best of our knowledge, there is no available experimental or theoretical optical data for the platinum nitride, and the present study may be the first one to calculate the optical spectra of platinum nitrides.
%
% =============================================================================================================
\section{Calculation Methods}\label{Calculation Methods}
% =============================================================================================================
%
% -------------------------------------------------------------------------------------------------------------
\subsection{\label{Stoichiometries and Crystal Structures}Stoichiometries and Crystal Structures}
% -------------------------------------------------------------------------------------------------------------
The structure and stability of solids are influenced by their chemical stoichiometry \cite{Crystal_Chemistry_of_Nitrides_review} and the electronic structure of the outer shells of atoms is a controlling factor in proposing any crystal structure \cite{Ag3N_structure_theory_1982}. Like other theoretical works (cf. Table \ref{platinum_nitrides_equilibrium_structural_properties}) we postulated various structure types that are not based on the observed fcc Pt sub-lattice. Our assignment of the following different chemical stoichiometries and crystal structures is based on the fact that many transition-metal nitrides (TMNs) are known to form more than one nitride \cite{StructuralInChem}.
Ni, which shares the same group with Pt, and Au which shares the same period with Pt in the periodic table, are known to form Ni$_3$N and Au$_3$N nitrides. Thus, it is of interest to consider more Pt atoms in the unit cell and less symmetric structures, and to know whether platinum can form Pt$_3$N with the reported structures of these and other 3:1 TMNs. In this work, the hypothetical Pt$_3$N is studied in the following seven structures:
 D0$_3$ (space group Fm$\bar{3}$m),
 A15 (space group Pm$\bar{3}$n),
 D0$_9$ (space group Pm$\bar{3}$m),
 L1$_2$ (space group Pm$\bar{3}$m),
 D0$_2$ (space group Im$\bar{3}$),
 $\epsilon$-Fe$_3$N (space group P6$_{3}$22)
 and RhF$_3$ (space group R$\bar{3}$c).

The following nine structures were assigned to PtN:
 B1 (space group Fm$\bar{3}$m),
 B2 (space group Pm$\bar{3}$m),
 B3 (space group F$\bar{4}3$m),
 B8$_{1}$ (space group P$6_{3}$/mmc),
 B$_{\text{k}}$ (space group P$6_{3}$/mmc),
 B$_{\text{h}}$ (space group P$\bar{6}$m$2$),
 B4 (space group P$6_{3}$mc),
 B17 (space group P$4_{2}$/mmc)
 and B24 (space group Fmmm).
 While 
C1 (space group Fm$\bar{3}$m),
 C2 (space group Pa$\bar{3}$),
 C18 (space group Pnnm)
 and CoSb$_{2}$ (space group P2$_1$/c)
 are the four structures which were proposed for PtN$_2$.
%
% ---------------------------------------------------------------------------------------------
\subsection{\label{Electronic Relaxation Details}Electronic Relaxation Details}
As implemented in the all-electron Vienna \textit{ab initio} Simulation Package (VASP) \cite{Vasp_ref_PhysRevB.47.558_1993,Vasp_ref_PhysRevB.49.14251_1994,Vasp_cite_Kressw_1996,
Vasp_PWs_Kresse_1996,DFT_VASP_Hafner_2008,PAW_Kresse_n_Joubert}, our electronic structure calculations were based on spin density functional theory (SDFT) \cite{SDFT_1972,SDFT_Pant_1972}. To solve the self-consistent Kohn-Sham (KS) equations \cite{KS_1965}
\begin{eqnarray}	\label{KS equations}
\begin{split}
  \Bigg \{ - \frac{\hbar^{2}} {2m_{e}}  \nabla^{2} &+ \int d\mathbf{r}^{\prime} \frac{n(\mathbf{r}^{\prime})}{|\mathbf{r}-\mathbf{r}^{\prime}|} + V_{ext}(\mathbf{r}) \quad \quad \quad \quad \quad 	\\
    &+ V_{XC}^{\sigma, \mathbf{k}}[n(\mathbf{r})] \Bigg \} \psi_{i}^{\sigma, \mathbf{k}}(\mathbf{r})  =  
   \epsilon_{i}^{\sigma, \mathbf{k}} \psi_{i}^{\sigma, \mathbf{k}}(\mathbf{r}),
\end{split}
\end{eqnarray}
where $i$, $\mathbf{k}$ and $\sigma$ are the band, $\mathbf{k}$-point and spin indices, respectively, the pseudo part of the KS orbitals $\psi_{i}^{\sigma , \mathbf{k}}(\mathbf{r})$ are expanded on plane-waves (PWs) basis. Only those PWs with kinetic energy $\frac{\hbar^{2}} {2m_{e}} |\mathbf{k} + \mathbf{G}| < E_{cut} = 600 \; eV$ were included. This always corresponds to changes in the total electronic energy and in the so-called Fermi energy $E_{F}$ that are less than $3 \; \text{m} eV/ \text{atom}$ and $1 \; \text{m}eV$, respectively.

For ionic relaxation, the Brillouin zones (BZs) were sampled using $13 \times 13 \times 13$ $\mathbf{\Gamma}$-centered Monkhorst-Pack meshes \cite{MP_k_mesh_1976}; while for the static calculations, $17 \times 17 \times 17$ meshes were used. Any increase in the density of this mesh corresponds to change in the total energy and in Fermi energy $E_{F}$ that are  less than $2 \; \text{m} eV/ \text{atom}$ and $0.02 \; eV$, respectively.

Partial occupancies were set using the tetrahedron method with Bl\"{o}chl corrections \cite{tetrahedron_method_theory_1971,tetrahedron_method_theory_1972,ISMEAR5_1994} for the static total energy and the electronic density of states (DOS) calculations; while in the ionic relaxation, the smearing method of Methfessel-Paxton (MP) \cite{MP_smearing_1989} was used. In the latter case, the smearing width was set such that the fictitious entropy is always less than $1 \; \text{m}eV/\text{atom}$.

The Perdew, Burke and Ernzerhof (PBE) \cite{PBE_GGA_1996,PBE_GGA_Erratum_1997,XC_PBE_1999} parametrization of the generalized gradient approximation (GGA) \cite{XC_GGA_1988,XC_GGA_applications_1992,
XC_GGA_applications_1992_ERRATUM} was employed for the exchange-correlation potentials $V_{XC}^{\sigma, \mathbf{k}}[n(\mathbf{r})]$. VASP treats the core-valence interactions, $V_{ext}(\mathbf{r})$ using the projector augmented wave (PAW) method \cite{PAW_Blochl, PAW_Kresse_n_Joubert}. The PAW potential explicitly treats the $2s^{2}2p^{3}$ electrons of N and the $5d^{9}6s^{1}$ electrons of Pt as valence electrons.

In the standard mode of VASP, while fully relativistic calculations are performed for the core-electrons, only scalar kinematic effects are incorporated to treat the valance electronic structure \cite{DFT_VASP_Hafner_2008}. It was found that this scheme is sufficient and the spin-orbit interactions have little effect on the macro-physical properties of platinum nitride \cite{PtN_n_PtN2_2005_July_comp_5+}. Thus, we made no effort to consider spin-orbit coupling of the valence electrons.

The relaxation of the electronic degrees of freedom was performed using the blocked Davidson iteration scheme \cite{Davidson_original_article_1975} as implemented in VASP. The electronic self-consistent (SC) convergence was considered to be achieved when the difference in the eigenvalues and in the total energy between two successive steps are both less than $1 \times 10^{-4} \; eV$.
%
% ----------------------------------------------------------------------------------------------------------
\subsection{\label{Geometry Optimization and EOS}Geometry Optimization and EOS} 
% ----------------------------------------------------------------------------------------------------------
At a set of isotropically varying volumes of the unit cells, ions with internal free parameters were relaxed until all Hellmann-Feynman force components \cite{Hellmann–Feynman_theorem} on each ion were less than $1 \times 10^{-2} \; eV/\text{\AA}$. Static total energy calculations (as described above) at each volume followed and the cohesive energy per atom was calculated from \cite{Grimvall,Suleiman_PhD_arxive_copper_nitrides_article}
\begin{eqnarray} \label{E_coh equation}
E_{coh}^{\text{Pt}_{m}\text{N}_{n}}  =   \frac{  E_{\text{solid}}^{\text{Pt}_{m}\text{N}_{n}} - Z \times \left( m E_{\text{atom}}^{\text{Pt}} + n E_{\text{atom}}^{\text{N}} \right) }{Z \times (m + n)}.
\end{eqnarray}
Here, $Z$ is the number of Pt$_{m}$N$_{n}$ per unit cell, $E_{\text{atom}}^{\text{Pt}}$ and $E_{\text{atom}}^{\text{N}}$ are the energies of the spin-polarized non-spherical isolated Pt and N atoms, $E_{\text{solid}}^{\text{Pt}_{m}\text{N}_{n}}$ are the bulk cohesive energies calculated by VASP with respect to spherical non spin-polarized reference atoms, and $m,n = 1,2 \text{ or } 3$ are the stoichiometric weights.

The calculated $E_{coh}$ per atom as a function of volume $V$ per atom were fitted to a Birch-Murnaghan 3rd-order equation of state (EOS)\cite{BM_3rd_eos}
\begin{eqnarray} \label{3rd_BM_eos}
\begin{split}
E(V) = E_{0} +  \frac{9V_{0}B_{0}}{16}  \left(    \left[ \left( \frac{V_{0}}{V} \right)^{\frac{2}{3} } - 1 \right]^{3} B_{0}^{\prime}   \quad   \quad   \quad \right.  \\ \left.    +  \left[ \left( \frac{V_{0}}{V} \right)^{\frac{2}{3} } - 1 \right]^{2} \left[6 - 4 \left( \frac{V_{0}}{V} \right)^{\frac{2}{3} }\right]  \right) \; .
\end{split}
\end{eqnarray}
The four equilibrium fitting parameters: the equilibrium volume $V_{0}$, the equilibrium cohesive energy $E_{0}$, the equilibrium bulk modulus
\begin{equation}	\label{B_0 eq}
B_{0} = -V \frac{\partial P}{\partial V}\Big|_{V=V_{0}} = -V \frac{\partial^{2} E}{\partial V^{2}}\Big|_{V=V_{0}}
\end{equation}
and its pressure derivative
\begin{equation}	\label{B^prime eq}
 B^{\prime}_{0} = \frac{\partial B}{\partial P} \Big|_{P=0} = \frac{1}{B_{0}} \left(  V \frac{\partial}{\partial V} (V \frac{\partial^{2} E}{\partial V^{2}}) \right)  \Big|_{V=V_{0}} 	
\end{equation}
were determined by a least-squares method.
%
% -----------------------------------------------------------------------------------------------------------
\subsection{Formation Energy}\label{Formation Energy}
% -----------------------------------------------------------------------------------------------------------
In addition to $E_{\text{coh}}$, another important measure of relative stability is the so-called formation energy $E_f$. Assuming that the solid Pt$_m$N$_n$ results from the interaction between the solid Pt(\textit{fcc}) metal and the gaseous N$_{2}$ through the chemical reaction
\begin{eqnarray} \label{E_f reaction}
m \text{Pt}^{\text{solid}} + \frac{n}{2} \text{N}_2^{\text{gas}} \rightleftharpoons \text{Pt}_m\text{N}_n^{\text{solid}},
\end{eqnarray}
$E_f$ can be obtained from
\begin{align} \label{formation energy equation}
E_f(\text{Pt}_m\text{N}_n^{\text{solid}}) =   E_\text{coh}(\text{Pt}_m\text{N}_n^{\text{solid}}) \quad \quad \quad \quad \quad \quad &
\nonumber \\
- \frac{  m E_\text{coh}(\text{Pt}^{\text{solid}}) + \frac{n}{2} E_\text{coh}(\text{N}_2^{\text{gas}})}{m + n} &		\; .
\end{align}
where $m,n=1,2,3$ are the stoichiometric weights and $E_\text{coh}(\text{Pt}_m\text{N}_n^{\text{solid}})$ is the cohesive energy per atom as obtained from Eq. \ref{E_coh equation}. The ground-state cohesive energy and other equilibrium properties of the elemental platinum $E_\text{coh}(\text{Pt}^{\text{solid}})$ in its \textit{fcc} A1 structure (space group Fm$\bar{3}$m No. 225) \cite{Wyckoff,Structure_of_Materials,Handbook_of_Mineralogy} are given in Table \ref{platinum_nitrides_equilibrium_structural_properties}.  We found the equilibrium cohesive energy of the molecular nitrogen ($E_\text{coh}(\text{N}_2^{\text{gas}})$) and its N--N bond length to be $-5.196 \; eV/\text{atom}$ and $1.113 \; \text{\AA}$. For details on how these properties were calculated, readers are referred to Ref. \onlinecite{Suleiman_PhD_arXiv2012_copper_nitrides_article}.
%
% -------------------------------------------------------------------------------------------------------------
\subsection{\label{GWA Calculations and Optical Properties}GWA Calculations and Optical Properties}
% -------------------------------------------------------------------------------------------------------------
Accurate quantitative description of optical properties of materials requires treatments beyond the level of DFT \cite{PAW_optics}. One choice is to follow the method which is provided by the many-body perturbation theory (MBPT).  In this approach one needs to solve a system of quasi-particle (QP) equations \cite{GWA_and_QP_review_1999,Kohanoff,JudithThesis2008}
\begin{eqnarray}	\label{QP equations}
\begin{split}
  \Bigg \{ - \frac{\hbar^{2}} {2m}  \nabla^{2} + & \int d\mathbf{r}^{\prime} \frac{n(\mathbf{r}^{\prime})}{|\mathbf{r}-\mathbf{r}^{\prime}|} + V_{ext}(\mathbf{r}) \Bigg \} \psi_{i,\mathbf{k}}^{QP}(\mathbf{r})  \\  +& \int d\mathbf{r}^{\prime} \Sigma(\mathbf{r},\mathbf{r}^{\prime};\epsilon_{i,\mathbf{k}}^{QP})  \psi_{i,\mathbf{k}}^{QP}(\mathbf{r}^{\prime}) = \epsilon_{i,\mathbf{k}}^{QP} \psi_{i,\mathbf{k}}^{QP}(\mathbf{r}).
\end{split}
\end{eqnarray}
In practice, one takes the wave functions $\psi_{i,\mathbf{k}}^{QP}(\mathbf{r})$ from the DFT calculations. However, this technique is computationally expensive, and we had to use less dense meshes of $\mathbf{k}$-points: ($10 \times 10 \times 10$) in the case of B17 and ($12 \times 12 \times 12$) in the case of B3. The quantity $\Sigma(\mathbf{r},\mathbf{r}^{\prime};\epsilon_{i,\mathbf{k}}^{QP})$ in Eqs. \ref{QP equations} above is known as self-energy. It contains all the static and dynamic exchange and correlation effects, including those neglected at the DFT-GGA level. When $\Sigma$ is written in terms of the Green's function $G$ and the frequency-dependent screened Coulomb interaction $W$ as
\begin{eqnarray}	\label{GW self-energy}
\begin{split}
\Sigma_{GW} = j \int d\epsilon^{\prime} G(\mathbf{r},\mathbf{r}^{\prime};\epsilon,\epsilon^{\prime}) W(\mathbf{r},\mathbf{r}^{\prime};\epsilon),
\end{split}
\end{eqnarray}
the approximation is referred to as $GW$ approximation. The dynamically screened interaction $W$ is related to the bare interaction $v$ via
\begin{eqnarray}
\begin{split}
W(\mathbf{r},\mathbf{r}^{\prime};\epsilon) = j \int d\mathbf{r}_{1} \varepsilon^{-1}(\mathbf{r},\mathbf{r}_{1};\epsilon)v(\mathbf{r}_{1},\mathbf{r}^{\prime}),
\end{split}
\end{eqnarray}
where the dielectric matrix $\varepsilon$ is calculated within the random phase approximation (RPA). The QP eigenvalues
\begin{eqnarray} \label{QP eigenvalues}
\begin{split}
\epsilon_{i,\mathbf{k}}^{QP} = \text{Re}  \left( 
\left\langle \psi_{i,\mathbf{k}}^{QP} \middle| 
H_{\text{KS}} - V_{XC} + \Sigma_{GW_{0}}
\middle|  \psi_{i,\mathbf{k}}^{QP} \right\rangle  	\right)
\end{split}
\end{eqnarray}
are updated in the calculations of $G$, while $W$ is kept at the DFT-RPA level. This is called the $GW_{0}$ self-consistent routine on $G$. After the execution of the fourth iteration, $\varepsilon$ is recalculated within the RPA using the updated QP eigenvalues \cite{Kohanoff,JudithThesis2008,VASPguide}. It is straightforward then to calculate all the frequency-dependent optical spectra (e. g.
 refractive index $n\left(\omega\right)$,
 extinction coefficient $\kappa\left(\omega\right)$,
 absorption coefficient $\alpha\left(\omega\right)$,
 reflectivity $R\left(\omega\right)$ and 
 transmitivity $T\left(\omega\right) = 1- R\left(\omega\right)$) from the real $\varepsilon_{\text{re}}(\omega)$ and the imaginary $\varepsilon_{\text{im}}(\omega)$ parts of $\varepsilon_{\text{RPA}}(\omega)$ \cite{Fox,dressel2002electrodynamics,Ch9_in_Handbook_of_Optics_2010}: 
\begin{align}
n\left(\omega\right) &= \frac{1}{\sqrt{2}} \left(  \left[  \varepsilon_{\text{re}}^{2}\left(\omega\right) + \varepsilon_{\text{im}}^{2}\left(\omega\right) \right]^{\frac{1}{2}} + \varepsilon_{\text{re}}\left(\omega\right) \right)^{\frac{1}{2}} 			\label{n(omega)}\\
\kappa\left(\omega\right) &= \frac{1}{\sqrt{2}} \left(  \left[  \varepsilon_{\text{re}}^{2}\left(\omega\right) + \varepsilon_{\text{im}}^{2}\left(\omega\right) \right]^{\frac{1}{2}} - \varepsilon_{\text{re}}\left(\omega\right) \right)^{\frac{1}{2}} 			\label{kappa(omega)}\\
\alpha\left(\omega\right) &= \sqrt{2} \omega  \left(  \left[  \varepsilon_{\text{re}}^{2}\left(\omega\right) + \varepsilon_{\text{im}}^{2}\left(\omega\right) \right]^{\frac{1}{2}}  - \varepsilon_{\text{re}}\left(\omega\right) \right)^{\frac{1}{2}} 		\label{alpha(omega)}\\
R\left(\omega\right) &= \left| \frac{\left[  \varepsilon_{\text{re}}\left(\omega\right) + j \varepsilon_{\text{im}}\left(\omega\right)  \right]^{\frac{1}{2}} - 1}{\left[  \varepsilon_{\text{re}}\left(\omega\right) + j \varepsilon_{\text{im}}\left(\omega\right)  \right]^{\frac{1}{2}} + 1} \right| ^{2}		\label{R(omega)}
\end{align}

It may be worth to emphasize here that for more accurate optical properties (e.g. more accurate amplitudes and positions of the characteristic peaks), electron-hole excitations should be calculated by solving the so-called Bethe-Salpeter equation, the equation of motion of the two-body Green function $G_2$. The latter can be evaluated on the basis of our obtained GW one-particle Green function $G$ and QP energies \cite{DFT_GW_BSE_Electron-hole_excitations_and_optical_spectra_from_first_principles_2000}.
%
% =============================================================================================================
\section{Results and Discussion}\label{Results and Discussion}
\begin{figure}[!]
%\begin{center}
% Figure environments same as 0.8 * \textwidth please
\includegraphics[width=0.45\textwidth]{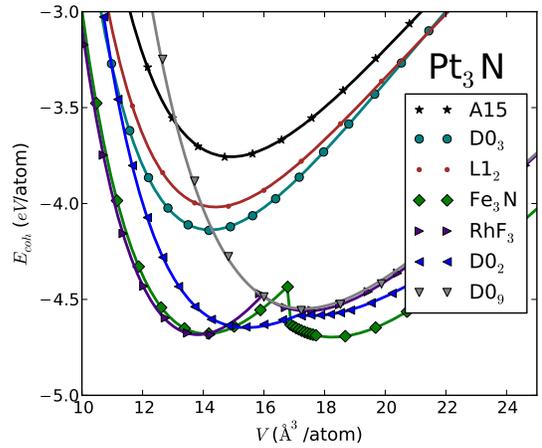}
\caption{\label{Pt3N1_ev_EOS}(Color online.) Cohesive energy $E_{\text{coh}} (eV/\text{atom})$ versus atomic volume $V$ (\AA$^{3}$/\text{atom}) for Pt$_3$N in seven different structural phases.}
%\end{center}
\end{figure}

\begin{figure}[!]
%\begin{center}
% Figure environments same as 0.8 * \textwidth please
\includegraphics[width=0.45\textwidth]{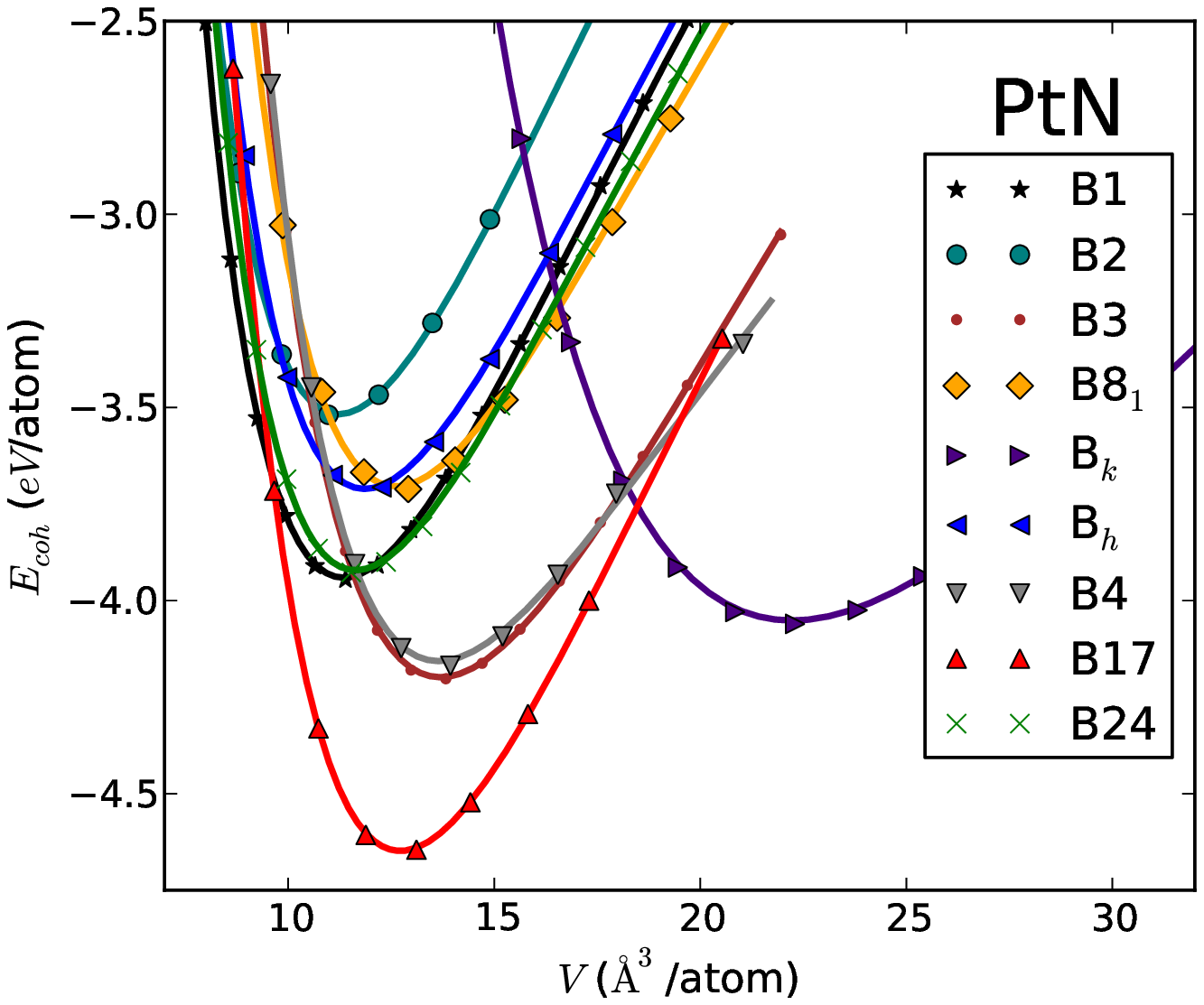}
\caption{\label{Pt1N1_ev_EOS}(Color online.) Cohesive energy $E_{\text{coh}} (eV/\text{atom})$ versus atomic volume $V$ (\AA$^{3}$/\text{atom}) for PtN in nine different structural phases.}
%\end{center}
\end{figure}

\begin{figure}[!]
%\begin{center}
% Figure environments same as 0.8 * \textwidth please
\includegraphics[width=0.45\textwidth]{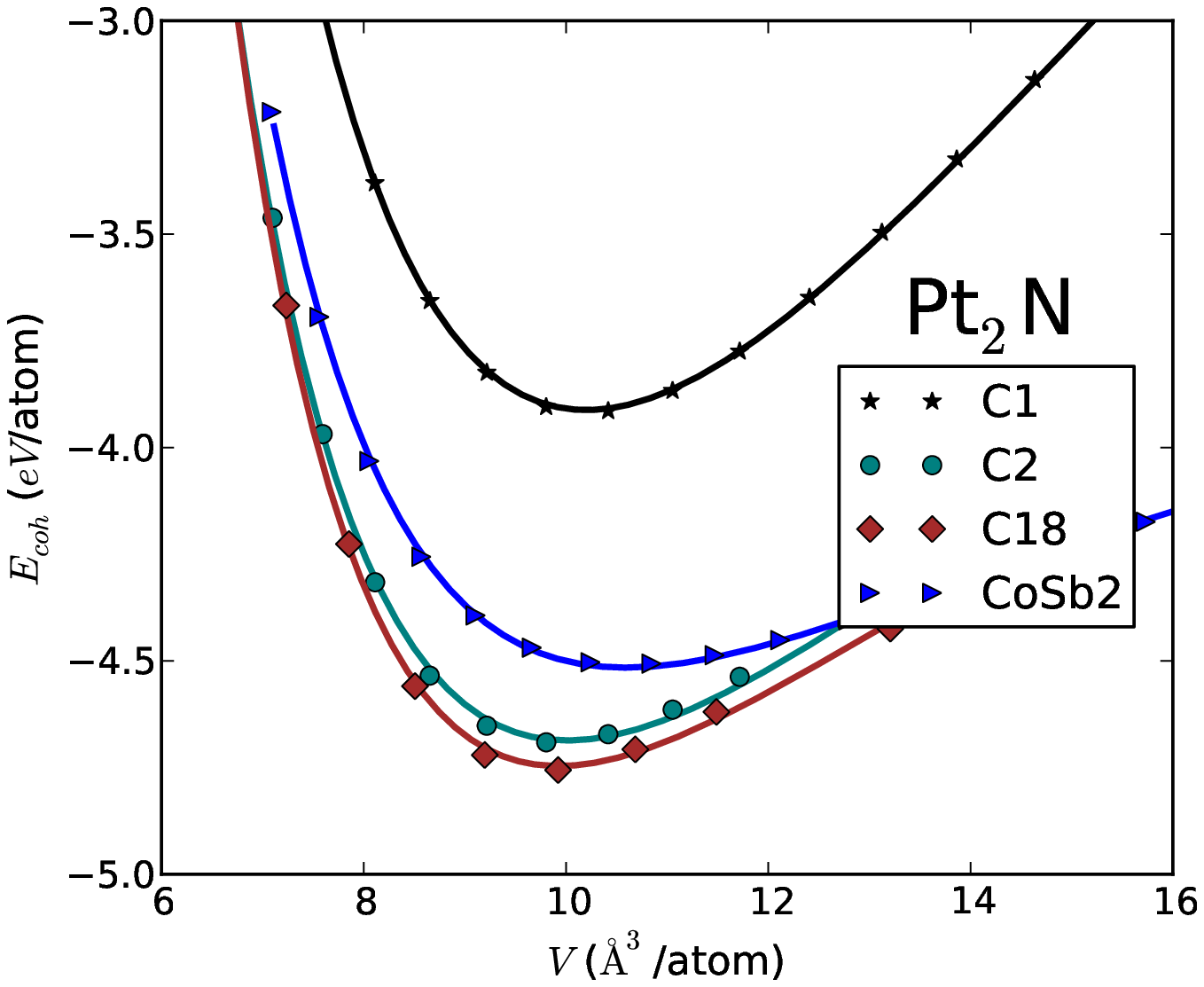}
\caption{\label{Pt1N2_ev_EOS}(Color online.) Cohesive energy $E_{coh} (eV/\text{atom})$ versus atomic volume $V (\AA^{3}/\text{atom})$ for PtN$_2$ in four different structural phases.}
%\end{center}
\end{figure}

\begin{figure*}[!]
%\begin{center}
% Figure environments same as 0.8 * \textwidth please
\includegraphics[width=0.95\textwidth]{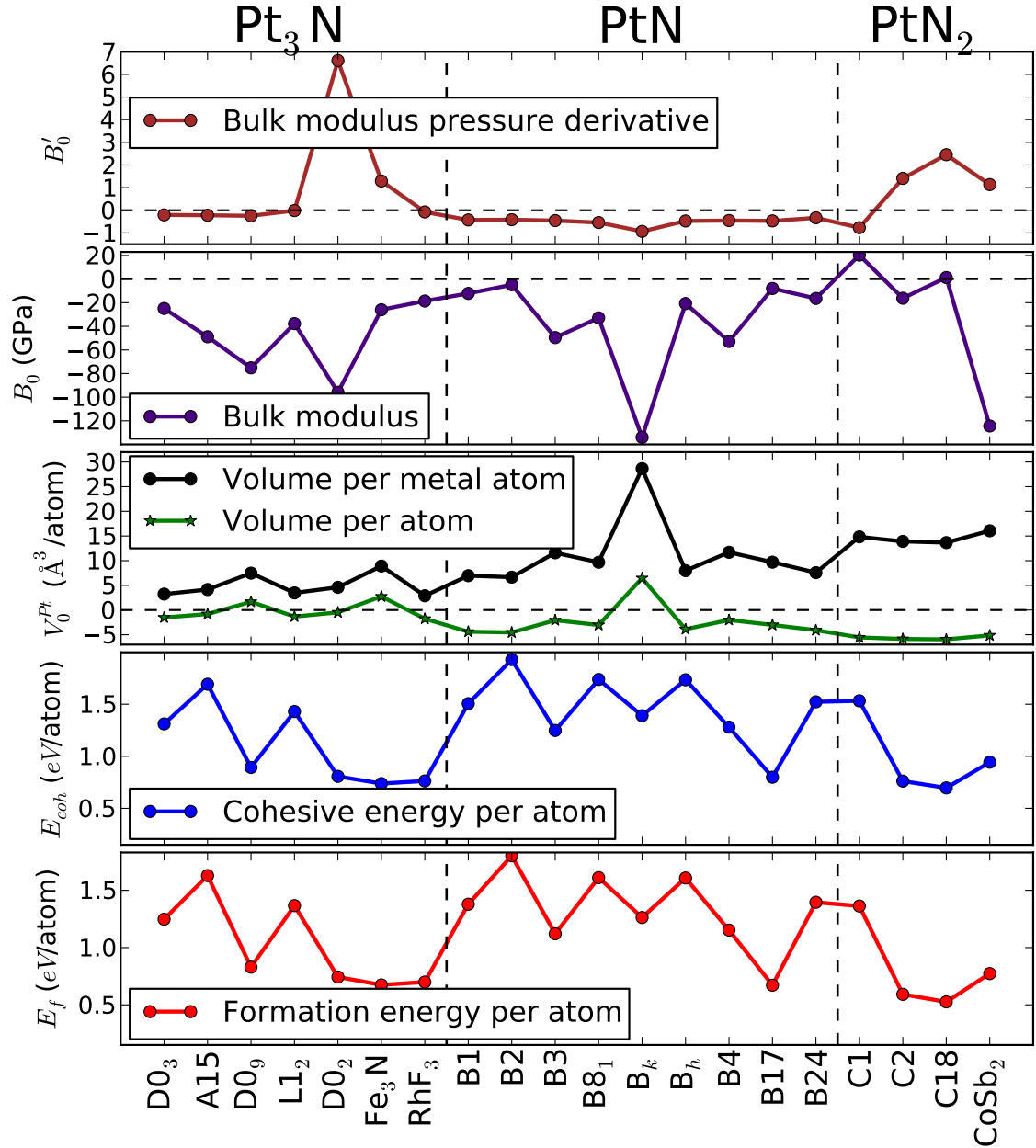}
\caption{\label{platinum_nitrides_equilibrium_properties}(Color online.) Calculated equilibrium properties of the twenty studied phases of platinum nitrides. All quantities are given relative to the corresponding ones of the \textit{fcc} crystalline elemental platinum given in the first row of Table \ref{platinum_nitrides_equilibrium_structural_properties}.}
%\end{center}
\end{figure*}

% =============================================================================================================
Cohesive energy $E_{\text{coh}}$ versus atomic volume $V_{0}$ equation of state (EOS) for the different phases of Pt$_3$N, PtN$_2$ and PtN are displayed graphically in Fig. \ref{Pt3N1_ev_EOS}, Fig. \ref{Pt1N1_ev_EOS} and Fig. \ref{Pt1N2_ev_EOS}, respectively. The corresponding obtained equilibrium structural parameters and energetic and elastic properties are presented in Table \ref{platinum_nitrides_equilibrium_structural_properties}. In this table, as well as in Fig. \ref{platinum_nitrides_equilibrium_properties}, structures are first grouped according to the nitrogen content, starting with the stoichiometry with the lowest nitrogen content Pt$_3$N, followed by the 1:1 series and ending with the nitrogen-richest PtN$_2$ group. Within each series, structures are ordered according to their structural symmetry, starting from the highest symmetry (i.e. the highest space group number) to the least symmetry. Whenever possible, our results are compared with experiment and with previous calculations. In the latter case, the calculations methods and the $XC$ functionals are indicated in the Table footnotes.

To study the effect of nitridation on the elemental Pt(A1) and to easly compare the properties of these phases relative to each other, the calculated equilibrium properties are displayed relative to the corresponding ones of Pt(A1) in Fig. \ref{platinum_nitrides_equilibrium_properties}.
%
% -------------------------------------------------------------------------------------------------
% -------------------------------------------------------------------------------------------------
%
% -------------------------------------------------------------------------------------------------
% -------------------------------------------------------------------------------------------------
%
% -------------------------------------------------------------------------------------------------
%
% -------------------------------------------------------------------------------------------------
%\begin{turnpage}
%\begingroup 
%\squeezetable 
% ===============================================================================================
\begin{table*} %\tiny
\caption{\label{platinum_nitrides_equilibrium_structural_properties}\small The calculated (\textit{Pres.}) zero-pressure properties of the platinum nitrides and the results of previous calculations (\textit{\textit{Comp.}}). The available experimental data are given in the last row.
}
\resizebox{1.0\textwidth}{!}{
%\begin{ruledtabular}
%\setlength{\extrarowheight}{3pt}
\begin{tabular}{lllllllllll}
%\begin{tabularx}{\textwidth}{XXXXXXXXXX}
\hline\noalign{\smallskip}
				&		& $a(\AA)$		   & $b(\AA)$			& $c(\AA)$		  & $\alpha(^{\circ})$ or $\beta(^{\circ})$		& $V_{0} (\text{\AA}^{3}/\text{atom})$   & $E_{\text{coh}}(eV/\text{atom})$	& $B_{0}(\text{GPa})$		& $B_{0}^{\prime}$ & $E_{f}(eV/\text{atom})$\\
\hline \hline % -------------------------------------------------------------------------------------
                		\multicolumn{11}{c}{\textbf{Pt}} \\
\hline % --------------------------------------------------------------------------------------------
\multirow{5}{*}{\textbf{A1}}        	&\textit{Pres.}	& $3.978$ & --    			& --  			   & --         		& $15.74$ & $-5.451$   & $242.999$ & $5.486$	&\\
              & \multirow{2}{*}{\textit{Expt.}}	&  ($3.9233 \pm 0.0007$)\footnotemark[1]   & --  & -- & --  & $15.097$\footnotemark[2]              & $-5.84$\footnotemark[3] & $278.3$\footnotemark[3], $280$\footnotemark[4]  & $5.18$\footnotemark[5] \\

              & 	&  $3.924$\footnotemark[31]   & --  & -- & --  &       & 	 & $249$\footnotemark[31]  & $5.23$\footnotemark[31] \\
              
             &\multirow{2}{*}{\textit{Comp.}}	& $3.90$\footnotemark[7], $3.890$\footnotemark[19], $3.981$\footnotemark[33],  & --	 & -- & --  &                     & $-7.04$\footnotemark[8], $-3.74$\footnotemark[9],    &$305$\footnotemark[7], $320$\footnotemark[19], $238$\footnotemark[20] & $5.16$\footnotemark[11], $5.30$\footnotemark[12], $5.25$\footnotemark[13]\\     

          &	& $3.967$\footnotemark[20], $3.966$\footnotemark[29]   & --  & -- & --  &  &  $-5.53$\footnotemark[10]  & $249$\footnotemark[29], $242$\footnotemark[30], $244.18$\footnotemark[33] & $5.23$\footnotemark[29], $5.83$\footnotemark[30], $5.7$\footnotemark[33] & \\

\hline % --------------------------------------------------------------------------------------------
                \multicolumn{11}{c}{\textbf{Pt$_3$N}} \\
\hline % --------------------------------------------------------------------------------------------
\multirow{1}{*}{\textbf{D0$_3$}}        &\textit{Pres.}	& $6.106$	&--                         &--                         &--                         & $14.23$ & $-4.140$ & $218.097$ & $5.282$ & $1.247$\\

\hline % --------------------------------------------------------------------------------------------
\multirow{1}{*}{\textbf{A15}}           &\textit{Pres.}	& $4.924$ &--                         &--                         &--                         & $14.92$ & $-3.759$ & $194.136$ & $5.266$ & $1.628$\\

\hline % --------------------------------------------------------------------------------------------
\multirow{1}{*}{\textbf{D0$_9$}}        &\textit{Pres.}	& $4.114$ &--                         &--                         &--                         & $17.41$ & $-4.558$ & $167.839$ & $5.241$ & $0.829$\\

\hline % --------------------------------------------------------------------------------------------
\multirow{1}{*}{\textbf{L1$_2$}}        &\textit{Pres.}	& $3.863$ &--                         &--                         &--                         & $14.41$ & $-4.021$ & $205.279$ & $5.472$ & $1.366$\\

\hline % --------------------------------------------------------------------------------------------
\multirow{1}{*}{\textbf{D0$_2$}}        &\textit{Pres.}	& $7.875$ &--                         &--                         &--                         & $15.26$ & $-4.644$ & $147.174$ & $12.098$ & $0.743$\\

\hline % --------------------------------------------------------------------------------------------
\multirow{1}{*}{\textbf{$\epsilon$-Fe$_3$N}}&\textit{Pres.}	& $5.680$ &--   & $5.293$ &--                         & $18.49$ & $-4.713$ & $217.035$ & $6.779$ & $0.674$ \\

\hline % --------------------------------------------------------------------------------------------
\multirow{1}{*}{\textbf{RhF$_3$}}       &\textit{Pres.}	& $5.463$ &--       &--    & $\alpha=58.640$ & $13.97$ & $-4.688$ & $224.419$ & $5.412$ & $0.699$ \\

\hline % --------------------------------------------------------------------------------------------
\multicolumn{11}{c}{\textbf{PtN}} \\
\hline % --------------------------------------------------------------------------------------------
\multirow{4}{*}{\textbf{B1}}   		&\textit{Pres.}	& $4.495$ & --                        & --	                   & --                        & $11.35$ & $-3.945$ & $230.869$	& $5.059$ & $1.378$ \\

	   &\multirow{3}{*}{\textit{Comp.}}	& $4.45$\footnotemark[24], $4.50$\footnotemark[25], $4.41$\footnotemark[26]      & --   & --	 &--   &  &                     &  $232$\footnotemark[24], $230$\footnotemark[25], $288$\footnotemark[26]        &           & \\     

	       		& 	& $4.471$\footnotemark[29]    & --	                & --	                    &--                       & $10.66$\footnotemark[32] &                     &  $251$\footnotemark[29], $242$\footnotemark[30], $294$\footnotemark[32]   & $4.00$\footnotemark[29], $4.78$\footnotemark[30]   & $1.365$\footnotemark[29]\\     

	       		& 	& $4.491$\footnotemark[33]    & --	                & --	                    &--                       &  &     &  $229.76$\footnotemark[33]   & $4.9$\footnotemark[33]   &	$[0.375 + E_f\text(B17)]$\footnotemark[34]	\\

\hline % --------------------------------------------------------------------------------------------
\multirow{2}{*}{\textbf{B2}}   		&\textit{Pres.}	& $2.819$ & --	                & --	                    &--                       & $11.20$ & $-3.522$ & $238.187$  & $5.070$  & $1.801$ \\

	       		& \textit{Comp.}	& $2.818$\footnotemark[33]    & --	                & --	                    &--                       &  &     &  $234.88$\footnotemark[33]   & $5.1$\footnotemark[33]   &		\\     

\hline % --------------------------------------------------------------------------------------------
\multirow{5}{*}{\textbf{B3}}   		&\textit{Pres.}	&  $4.782$ & --	                & --	                    &--                       & $13.67$ & $-4.203$ & $193.466$ & $5.031$ & $1.120$ \\
       		&\multirow{4}{*}{\textit{Comp.}}	& $4.7217$\footnotemark[14], $4.8250$\footnotemark[15], $4.779$\footnotemark[33],      & --	                & --	                    &--                       &                     &                     &  $243.3$\footnotemark[14], $196.3$\footnotemark[15], $190.61$\footnotemark[33],        &            $5.1$\footnotemark[33], & \\     

                               		&	& $4.6833$\footnotemark[16], $4.7889$\footnotemark[17], $4.8114$\footnotemark[18],      & --	                & --	                    &--                       &                     &                     &  $271.9$\footnotemark[16], $192.7$\footnotemark[17], $184$\footnotemark[18],        &           & $0.95$\footnotemark[27] \\     

                               		&	& $4.692$\footnotemark[19], $4.780$\footnotemark[20], $4.760$\footnotemark[29],    & --	                & --	&--   &   &  &  $244$\footnotemark[19], $194$\footnotemark[20], $213$\footnotemark[29], & $4.00$\footnotemark[29]     & $1.1$\footnotemark[29]\\     

                       		&	& $4.80$\footnotemark[24]\textsuperscript{,}\footnotemark[25], $4.70$\footnotemark[26], $4.699$\footnotemark[34]      & --	                & --	                    &--                       &                     &                     &  $192$\footnotemark[24], $190$\footnotemark[25], $232$\footnotemark[26], $217$\footnotemark[30]        & $3.62$\footnotemark[30]  & $[0.21 + E_f\text(B17)]$\footnotemark[34]\\     

\hline % --------------------------------------------------------------------------------------------
\multirow{1}{*}{\textbf{B8$_{1}$}}   	&\textit{Pres.}	&  $3.482$ &--  		        & $4.843$ &--                       & $12.71$ & $-3.713$ & $210.165$ & $4.945$ & $1.610$ \\

\hline % --------------------------------------------------------------------------------------------
\multirow{1}{*}{\textbf{B$_{\text{k}}$}}&\textit{Pres.}	&  $3.378$ &--                         & $8.986$                    &--                       & $22.20$  & $-4.061$ & $108.968$ & $4.553$ & $1.262$ \\

\hline % --------------------------------------------------------------------------------------------
\multirow{1}{*}{\textbf{B$_{\text{h}}$}}&\textit{Pres.}	&  $3.039$ &--                         & $2.966$                    &--                       & $11.86$ & $-3.716$ & $222.279$ & $5.014$ & $1.607$ \\

\hline % --------------------------------------------------------------------------------------------
\multirow{2}{*}{\textbf{B4}}   		&\textit{Pres.}	&  $3.382$ &--   & $5.539$ &--                       & $13.72$ & $-4.171$ & $190.130$ & $5.033$ & $1.152$ \\

	       		& \textit{Comp.}	& $3.386$\footnotemark[33]    & --	 & $5.529$\footnotemark[33]	                    &--                       &  &     &  $191.06$\footnotemark[33]   & $4.7$\footnotemark[33]   &		\\     

\hline % --------------------------------------------------------------------------------------------
\multirow{2}{*}{\textbf{B17}}   	&\textit{Pres.}	&	$3.069$ &--     & $5.403$ &--                       & $12.72$ & $-4.652$ & $235.041$ & $5.018$  & $0.671$ \\
        	&	\textit{Comp.}	& $3.323$\footnotemark[34] &--  & $4.579$\footnotemark[34]  &--                       &                     &                     &          &        &   \\ 

\hline % --------------------------------------------------------------------------------------------
\multirow{2}{*}{\textbf{B24}}   	&\textit{Pres.}	& $4.216$ & $4.472$ & $4.948$ &--                       & $11.66$ & $-3.928$ & $ 226.608$ & $5.153$ & $1.395$ \\

       	&	\textit{Comp.}$^{*}$	& $3.972$\footnotemark[34] & $3.977$\footnotemark[34] & $6.022$\footnotemark[34]  &--     &    &   & $270$\footnotemark[34] &     & $[0.085 + E_f\text(B17)]$\footnotemark[34] \\ 

\hline % --------------------------------------------------------------------------------------------
		\multicolumn{11}{c}{\textbf{PtN$_2$}} \\
\hline % --------------------------------------------------------------------------------------------
\multirow{3}{*}{\textbf{C1}}        &\textit{Pres.}	& $4.963$ &--                         &--                         &--                         & $10.19$ & $-3.918$ & $263.295$ & $4.717$ & $1.363$ \\
                &\multirow{2}{*}{\textit{Comp.}}	& $4.9428$\footnotemark[14], $5.0403$\footnotemark[15],  &--                         &--                         &--                         &                     &                     & $322.1$\footnotemark[14], $267.2$\footnotemark[15], $269$\footnotemark[29],   & $4.00$\footnotemark[29] & $1.167$\footnotemark[27] \\

                               		&	& $4.866$\footnotemark[19], $4.958$\footnotemark[20], $4.939$\footnotemark[30]  & --	                & --	&--   &   &  &  $316$\footnotemark[19], $264$\footnotemark[20], $260$\footnotemark[30]  & $4.73$\footnotemark[30] & $1.317$\footnotemark[30] \\

\hline % --------------------------------------------------------------------------------------------
\multirow{3}{*}{\textbf{C2}}            &\textit{Pres.}	& $4.912$ &--                         &--                         &--                         & $9.882$ & $-4.689$ & $ 226.779$ & $6.893$ & $0.592$ \\

        & \multirow{2}{*}{\textit{Comp.}}	& $4.87$\footnotemark[27] &--                         &--         &--   & $9.12$\footnotemark[32]  &     &          &           & $0.267$\footnotemark[27], $0.24$\footnotemark[28]\\

        & 	& $4.848$\footnotemark[29], $4.874$\footnotemark[35] &--     &--   &-- & $9.65$\footnotemark[35]    &          & $305$\footnotemark[29], $285$\footnotemark[30], $300$\footnotemark[35]          & $4.00$\footnotemark[29], $5.50$\footnotemark[30] & $0.64$\footnotemark[29], $0.212$\footnotemark[35]\\

\hline % --------------------------------------------------------------------------------------------
\multirow{2}{*}{\textbf{C18}}           &\textit{Pres.}	& $3.036$ & $3.984$ & $4.862$ &--                         & $9.800$ & $-4.755$ & $244.320$ & $7.938$ & $0.526$ \\

			&\multirow{1}{*}{\textit{Comp.}}		& $3.778$\footnotemark[35]   & $4.880$\footnotemark[35]   & $3.208$\footnotemark[35] &--    & $9.827$\footnotemark[35] &   & $286$\footnotemark[35]  &  & $0.249$\footnotemark[35] \\

\hline % --------------------------------------------------------------------------------------------
\multirow{2}{*}{\textbf{CoSb$_2$}}      &\textit{Pres.}	& $5.460$ & $5.163$ & $9.374$ & $\beta=151.225$                    & $10.60$ & $-4.508$ & $118.594$ & $6.619$ & $0.773$ \\

			&\multirow{1}{*}{\textit{Comp.}}		& $4.950$\footnotemark[35]   & $4.880$\footnotemark[35]   & $4.950$\footnotemark[35] & $99.50$\footnotemark[35]   & $9.827$\footnotemark[35] &   & $289$\footnotemark[35]  &  &  $0.248$\footnotemark[35] \\

\hline % --------------------------------------------------------------------------------------------
                \multicolumn{11}{c}{\textbf{Experiment}} \\
\hline % --------------------------------------------------------------------------------------------
                         & 	& $(4.8032 \pm 5)$\footnotemark[21]\textsuperscript{,}\footnotemark[22], $4.8041(2)$\footnotemark[23] &  --   & --         & --                &                     &                     & $(372 \pm 5)$\footnotemark[21], $(354 \pm 5)$\footnotemark[22]  & $4.0$\footnotemark[21], $5.26$\footnotemark[22] &	\\
\hline \hline	% -----------------------------------------------------------------------------------
\end{tabular}
%\end{tabularx}
%\end{ruledtabular}
}	% for the \resizebox command

\footnotetext[1]{Ref. \onlinecite{Jerry_1974}: This is an average of 23 experimental values, at room temperature.}
\footnotetext[2]{Ref. \onlinecite{Jerry_1974}: at room temperature.}
\footnotetext[3]{Ref. \onlinecite{Kittel}: Cohesive energies are given at $0 \; K$ and $1 \text{ atm} = 0.00010 \; \text{GPa}$; while bulk moduli are given at room temperature.}
\footnotetext[4]{Ref. (25) in \onlinecite{B_prime_1997_theory_comp_n_exp}: at room temperature.}
\footnotetext[5]{See Refs. (8)--(11) in \onlinecite{B_prime_1997_theory_comp_n_exp}.}
\footnotetext[7]{Ref. \onlinecite{elemental_metals_1996_comp}: using the full-potential linearized augmented plane waves (LAPW) method within LDA.}
\footnotetext[8]{Ref. \onlinecite{elemental_metals_2008_comp}: using the projector augmented wave (PAW) method within LDA.}
\footnotetext[9]{Ref. \onlinecite{elemental_metals_2008_comp}: using the projector augmented wave (PAW) method within GGA(PW91).}
\footnotetext[10]{Ref. \onlinecite{elemental_metals_2008_comp}: using the projector augmented wave (PAW) method within GGA(PBE).}
\footnotetext[11]{Ref. \onlinecite{B_prime_1997_theory_comp_n_exp}: using the so-called method of transition metal pseudopotential theory; a modified form of a method proposed by Wills and Harrison to represent the effective interatomic interaction.}
\footnotetext[12]{Ref. \onlinecite{B_prime_1997_theory_comp_n_exp}: using a semi-empirical estimate based on the calculation of the slope of the shock velocity \textit{vs.} particle velocity curves obtained from the dynamic high-pressure experiments.  The given values are estimated at $\sim 298 \; K$.}
\footnotetext[13]{Ref. \onlinecite{B_prime_1997_theory_comp_n_exp}: using a semi-empirical method in which the experimental static $P-V$ data are fitted to an EOS form. The given values are estimated at $\sim 298 \; K$.}

\footnotetext[14]{Ref. \onlinecite{PtN_n_PtN2_2005_July_comp_5+}: using the ultrasoft pseudopotential (USPP) method within LDA. $B_0$'s are calculated from elastic constants.}
\footnotetext[15]{Ref. \onlinecite{PtN_n_PtN2_2005_July_comp_5+}: using the ultrasoft pseudopotential (USPP) method within GGA. $B_0$'s are calculated from elastic constants.}
\footnotetext[16]{Ref. \onlinecite{PtN_n_PtN2_2005_July_comp_5+}: using the projector augmented wave (PAW) method within LDA. $B_{0}^{\prime}$ is set to be $4$.}
\footnotetext[17]{Ref. \onlinecite{PtN_n_PtN2_2005_July_comp_5+}: using the projector augmented wave (PAW) method within GGA. $B_{0}^{\prime}$ is set to be $4$.}
\footnotetext[18]{Ref. \onlinecite{PtN_n_PtN2_2005_July_comp_5+}: using fully relativistic full-potential linearized augmented plane waves (LAPW) method within GGA.}

\footnotetext[19]{Ref. \onlinecite{PtN_PtN2_2005_comp}: using the full-potential linearized augmented plane waves (LAPW) method within LDA.}
\footnotetext[20]{Ref. \onlinecite{PtN_PtN2_2005_comp}: using the full-potential linearized augmented plane waves (LAPW) method within GGA(PBE).}

\footnotetext[21]{Ref. \onlinecite{first_PtN_2004_exp}: The experimental evolution of the volume with pressure was fitted with a Birch-Murnaghan EOS, but $B_{0}^{\prime}$ was set to be $4$.}
\footnotetext[22]{Ref. \onlinecite{first_PtN_2004_exp}: The experimental evolution of the volume with pressure was fitted with a Birch-Murnaghan EOS, but $B_{0,\text{Pt}}^{\prime} = 5.26$ was fixed.}
\footnotetext[23]{Ref. \onlinecite{first_PtN_2004_exp}: From XRD measurements at $0.1 \; \text{MPa}$.}

\footnotetext[24]{Ref. \onlinecite{Mysterious_Platinum_Nitride_2006_comp}: using the full-potential linearized augmented plane waves (FPLAPW) method within GGA(PBE).}
\footnotetext[25]{Ref. \onlinecite{Mysterious_Platinum_Nitride_2006_comp}: using pseudopotentials method within GGA(PBE).}
\footnotetext[26]{Ref. \onlinecite{Mysterious_Platinum_Nitride_2006_comp}: using pseudopotentials method within LDA.}

\footnotetext[27]{Ref. \onlinecite{PtN_n_PtN2_2006_exp_n_comp}: using the PAW method within GGA(PW91), but the experimental value of $E_\text{coh}(\text{N}_2^{\text{gas}})$ in Eq. \ref{formation energy equation} was used.}
\footnotetext[28]{Ref. \onlinecite{PtN_n_PtN2_2006_exp_n_comp}: using the full-potential linear-augmented plane-wave method.} 

\footnotetext[29]{Ref. \onlinecite{positive_Ef_n_PtN2_2006_comp_Scandolo}: using pseudopotentials method within GGA(PBE).}
\footnotetext[30]{Ref. \onlinecite{positive_Ef_n_PtN2_2006_comp_Scandolo}: using pseudopotentials method within GGA(PBE).}

\footnotetext[31]{Ref. \onlinecite{PhysRevB.70.094112}.}

\footnotetext[32]{Ref. \onlinecite{PtN_PtN2_2009_comp}: using the pseudopotential method within LDA.}

\footnotetext[33]{Ref. \onlinecite{PtN_2007_comp}: using the full potential augmented plane wave plus
local orbitals (APW+lo) method within GGA(PBE).}

\footnotetext[34]{Ref. \onlinecite{PtN_2006_comp}: using the Vanderbilt ultrasoft pseudopotentials within LDA. $^{*}$ The data from Ref. \onlinecite{PtN_2006_comp} are for a face-centered orthorhombic structure (space group Fddd) which is not the same as our studied face-centered orthorhombic B24 structure (space group Fmmm).}

\footnotetext[35]{Ref. \onlinecite{AuN2_and_PtN2_2010_comp}: using Vanderbilt USPPs within GGA(PBE). $B_0$'s are calculated from the elastic constants. $E_\text{coh}(\text{N}_2^{\text{solid}})$ was used in Eq. \ref{formation energy equation} instead of $E_\text{coh}(\text{N}_2^{\text{gas}})$.}

\end{table*}
%\end{turnpage}
%\endgroup
% -------------------------------------------------------------------------------------------------
%
% -------------------------------------------------------------------------------------------------
% -------------------------------------------------------------------------------------------------
%
% -----------------------------------------------------------------------------------------------------------
\subsection{EOS and Relative Stabilities}\label{EOS and Relative Stabilities}
% -----------------------------------------------------------------------------------------------------------
Fig. \ref{Pt3N1_ev_EOS} reveals that Pt$_3$N in its least symmetric phase, the trigonal (rhombohedric) structure of RhF$_3$, is the most favorable phase in this series. However, after $\sim 15.9 \; \text{\AA}^{3}/\text{atom}$ the EOS of Pt$_3$N(RhF$_3$) is almost identical with the EOS of Pt$_3$N in the simple cubic structure of the anti-ReO$_3$ (D0$_9$). They share a minimum at $\sim (17.4 \; \text{\AA}^{3}/\text{atom}, -4.56 \; eV)$. Very close to this point, at $\sim (17.23 \; \text{\AA}^{3}/\text{atom}, -4.59 \; eV)$, the EOS of Pt$_3$N(D0$_2$) has a kink due to a change in the positions of some Pt ions. % from 48h to 24g Wyckoff positions.

The EOS of Pt$_3$N(Fe$_3$N) has two minima located at $(14.11 \; \text{\AA}^{3}/\text{atom}, -4.697 \; eV)$ and $(18.26 \; \text{\AA}^{3}/\text{atom}, -4.679 \; eV)$. Thus, the two minima are very close in energy but, due to the difference in $V_0$, they correspond to bulk moduli of $222.7 \; \text{GP}$ and $169.0 \; \text{GP}$, respectively. The Pt ions are in the $6g$ Wyckoff positions: $(x,0,0), (0,x,0), (-x,-x,0), (-x,0,\frac{1}{2}), (0,-x,\frac{1}{2})$ and $(x,x,\frac{1}{2})$. Upon ion relaxation of Pt$_3$N(Fe$_3$N), atomic positions change from $x \sim \frac{1}{3}$ to $x=\frac{1}{2}$ causing the sudden change in the potential surface (at $\sim 16.83 \; \text{\AA}^3/\text{atom}$) as the bulk Pt$_3$N(Fe$_3$N) being decompressed (Fig. \ref{Pt3N1_ev_EOS}). It may be worth mentioning here that Ag$_3$N(Fe$_3$N) \cite{Suleiman_PhD_arXiv2012_silver_nitrides_article} and Cu$_3$N(Fe$_3$N) \cite{Suleiman_PhD_arXiv2012_copper_nitrides_article} were found to behave in a similar manner.

Hence, one of the two minima in the EOS of Pt$_3$N(RhF$_3$) is shared with the minimum of the EOS of Pt$_3$N(D0$_9$) and the other is shared with one of the two minima of Pt$_3$N(Fe$_3$N).

The crossings of the less stable D0$_3$, L1$_2$ and A15  EOS curves with the more stable D0$_9$ EOS at the left side of their equilibria indicates that D0$_9$ would not survive under pressure and that possible pressure-induces phase transitions from the latter phase to the former ones may occur.

Fig. \ref{platinum_nitrides_equilibrium_properties} shows that the Pt$_3$N most stable phases may energetically compete with the PtN and PtN$_2$ most stable ones. However, from the foregoing discussion, it seems that Pt$_3$N would not have a simple potential surface.\\

Using the full potential augmented plane wave plus local orbitals (APW+lo) method within GGA(PBE), the energy-volume EOS's for B1, B2, B3 and B4 have been studied by the authors of Ref. \onlinecite{PtN_2007_comp}. Some of their obtained equilibrium properties are included and referred to in Table \ref{platinum_nitrides_equilibrium_structural_properties}. Within the considered parameter sub-space, our obtained EOS's (Fig. \ref{Pt1N1_ev_EOS}), relative stabilities, and equilibrium structural parameters and mechanical properties (Table \ref{platinum_nitrides_equilibrium_structural_properties}) are in excellent agreement with their findings. However, relaxing the $c/a$ parameter, they obtained an additional EOS which lies below all the other considered ones, but its equilibrium $B_0$ is significantly smaller.

From Fig. \ref{Pt1N1_ev_EOS}, it is evident that PtN(B17) is the energetically most stable phase in the PtN series. The difference in the equilibrium $E_{\text{coh}}$ between PtN(B17) and the next (less) stable phase, PtN(B3), is about $0.5 \; eV$ (Table \ref{platinum_nitrides_equilibrium_structural_properties}). This difference was found by other researchers \cite{Mysterious_Platinum_Nitride_2006_comp} to be $0.9 - 1.05 \; eV$. The crossings of the EOS curve of B17 with some of those of less stability at the left side of their equilibria reveals possible pressure-induces phase transitions. To closely investigate these transitions, we plot the corresponding relations between enthalpy $H = E(V) + PV$ and the imposed external pressure $P$. Possible transitions and the pressures at which they occur are carefully depicted. A point where two $H(P)$ curves (of two modifications with the same chemical stoichiometry \cite{PtN_PtN2_2009_comp}) meet represents a phase transition from the phase with the higher $H$ to the one with the lower $H$ \cite{Grimvall}. From the $H(P)$ diagrams (not shown here) we found that PtN(B17) would transform to PtN(B1, B2, B$_h$ or B24) at $\sim 93 \; \text{GPa}$, $\sim 143 \; \text{GPa}$, $\sim 193 \; \text{GPa}$ or $\sim 123 \; \text{GPa}$, respectively.

\begin{figure}[!]
%\begin{center}
% Figure environments same as 0.8 * \textwidth please
\includegraphics[width=0.45\textwidth]{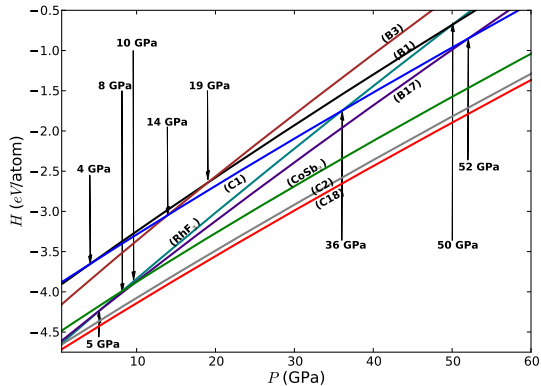}
\caption{\label{mixture_PtmNn_hp_EOS}(Color online.)
Enthalpy $H$ \textit{vs.} pressure $P$ equation of state (EOS) for the most favorable Pt$_3$N phase (RhF$_3$), the three proposed PtN modifications (B1, B3 and B17), and the four considered PtN$_2$ structures in the present work (C1, C2, C18 and CoSb$_2$). The arrows indicate the pressures at which curves cross each other.}
%\end{center}
\end{figure}

It may be worth to mention here a few points about this B17 structure:
 (i) It was theoretically predicted to be the ground-state structure of CuN \cite{Suleiman_PhD_arXiv2012_copper_nitrides_article}, AgN \cite{Suleiman_PhD_arXiv2012_silver_nitrides_article}, AuN \cite{Suleiman_PhD_SAIP2012_gold_nitrides_article} and PdN \cite{Suleiman_PhD_SAIP2012_palladium_nitrides_article}.
 (ii) The same foregoing phase PtN-PtN structural pressure-induced transitions have been predicted for PdN, but at relatively smaller pressures in the range $(25.8 \sim 62.1 \; \text{GPa})$ \cite{Suleiman_PhD_SAIP2012_palladium_nitrides_article}.
 (iii) B17 is the structure of PtS \cite{B17_structure_2} and PtO \cite{Mysterious_Platinum_Nitride_2006_comp}.
 (iv) It was found by other authors to be a possible ground state for PtN \cite{Mysterious_Platinum_Nitride_2006_comp}.
 (v) The B17 structure has an fcc Pt sub-lattice (as the synthesized platinum nitride), but it is tetragonal and the sub-lattice are highly distorted ($c/a \approx \sqrt{3}$ versus $c/a = \sqrt{2}$ for ideal fcc), and probably because of this distortion it was rejected by the platinum nitride synthesizers \cite{first_PtN_2004_exp}.
 (vi) Fig. \ref{mixture_PtmNn_hp_EOS} that B17 is energetically favorable over B1 and B3 at all pressures.

Nevertheless, PtN(B17) was found to be elastically unstable \cite{PtN_2006_comp}.
%Either: the most stable phase is tetragonal, but not necessarily B17. Or the tetragonal phase might be this B17; but an intermediate phase may exist. However, the latter is surely not among the modifications we have considered, and we could not figure out the series of the phase transitions from B3 to B17. OR: The most stable phase is tetragonal, but not necessarily B17.

Assuming 1:1 stoichiometry, the first platinum nitride synthesizers assigned the B3 structure for their product \cite{first_PtN_2004_exp}. However, it was shown in the same work that PtN(B3) should break down or transform at pressures above $12 \; \text{GPa}$. In agreement with this experimental prediction, Fig. \ref{mixture_PtmNn_hp_EOS} shows that PtN(B3) would not survive at pressures above $19 \; \text{GPa}$ where the B3$\rightarrow$B1 phase transition occurs. Other theoretical works also predicted that B1 becomes more favorable than B3 structure above
 $13.3 \; \text{GPa}$\cite{Mysterious_Platinum_Nitride_2006_comp},
 $\sim 15 \; \text{GPa}$\cite{PtN_PtN2_2009_comp},
 $16.5 \; \text{GPa}$\cite{Mysterious_Platinum_Nitride_2006_comp}, and 
 $17.6 \; \text{GPa}$\cite{Mysterious_Platinum_Nitride_2006_comp}.

% -------------------------------------------------------------------------------------------------
% -------------------------------------------------------------------------------------------------

Therefore, we support Ref. \onlinecite{Mysterious_Platinum_Nitride_2006_comp} on the judgment that, unless the PtN(B3) was formed upon depressurization, its production at $45-50 \; \text{GPa}$ \cite{first_PtN_2004_exp} is questioned. Further, first-principles calculations showed that PtN(B3) is elastically unstable \cite{PtN_n_PtN2_2005_July_comp_5+,PtN_PtN2_2005_comp,positive_Ef_n_PtN2_2006_comp_Scandolo}, and that it may distort spontaneously to a tetragonal lattice to lower the energy \cite{PtN_PtN2_2005_comp}.\\

In the PtN$_2$ series, we can see from Table \ref{platinum_nitrides_equilibrium_structural_properties} and from Fig. \ref{platinum_nitrides_equilibrium_properties} that PtN$_2$ in the simple orthorhombic structure of FeS$_{2}$ marcasite (C18) is the most stable phase, while the face-centered cubic structure of CaF$_{2}$ fluorite (C1) is significantly the least favorable structure. Yet, Fig. \ref{mixture_PtmNn_hp_EOS} reveals that the latter PtN$_2$(C1) is more favorable than the proposed PtN(B1, B3 and B17) at pressures above $4 \; \text{GPa}$, $14 \; \text{GPa}$ and $52 \; \text{GPa}$, respectively. Others \cite{Mysterious_Platinum_Nitride_2006_comp} found PtN$_2$(C1) to be more favorable than PtN(B3) at pressures above $30 \; \text{GPa}$.

In contrast to our enthalpy-pressure EOS's in Fig. \ref{mixture_PtmNn_hp_EOS}, Chen, Tse and Jiang \cite{AuN2_and_PtN2_2010_comp} obtained an $H(P)$ curve for C18 which lies always above the curve for C2 and coincides with the one of CoSb$_2$. They concluded that C2 is the most stable structure among these three modifications. While we sticked to the original C18 relative dimensions, it seems that Chen, Tse and Jiang tried to optimize the lattice parameters ratios (see Table \ref{platinum_nitrides_equilibrium_structural_properties}). However, the $c:a:b$ ratio they obtained is very close to our $a:b:c$ ratio, and the difference in $V_0$ is less than $0.03 \; \text{\AA}^{3}/\text{atom})$\footnote{Surprisingly, Chen, Tse and Jiang \cite{AuN2_and_PtN2_2010_comp} got exactly the same $V_0$ values for C2 and C18 within both GGA and LDA; but the average values they gave are different! Thus, we suspect the equal $V_0$ values they gave for C2 and C18 in both GGA and LDA (see Table 1 in that article); and it may be a typo.}. Another difference is the atomic electronic configuration of Pt $5d^86s^2$ they used. Nevertheless, they agreed with us that in the $0 - 60 \; \text{GPa}$ pressure range, no transition between these three phases occurs.\\

Comparing the relative stability of the three most stable compositions, we find from Table \ref{platinum_nitrides_equilibrium_structural_properties} and from Fig. \ref{platinum_nitrides_equilibrium_properties} that PtN$_2$(C18) is the most favorable, followed by Pt$_3$N(RhF$_3$), and the least stable phase is PtN(B17). However, the differences in their equilibrium $E_{\text{coh}}$ lies within a narrow range of $0.036 \; \text{eV}$. Relative to their parent metal, all phases have higher $E_{\text{coh}}$, i.e. they are less bound than Pt(A1). Hence, we found, as other theoretical works \cite{PtN_PtN2_2005_comp}, that platinum nitride can be stabilized in stoichiometries and structures other than that proposed by the first synthesizers \cite{first_PtN_2004_exp}.

In Ref. \onlinecite{Mysterious_Platinum_Nitride_2006_comp}, the energy-volume EOS for B1, B3, B17, C1, and C2 have been studied using DFT-GGA. Within this parameter sub-space, our obtained EOS's (Figs. \ref{Pt1N1_ev_EOS}, \ref{Pt1N2_ev_EOS} and \ref{platinum_nitrides_equilibrium_properties}) are in excellent agreement with the findings of \onlinecite{Mysterious_Platinum_Nitride_2006_comp}. From the relative enthalpy-pressure diagrams \footnote{These are $H(P)$ diagrams but relative to their elemental constituents.}, Ref. \onlinecite{Mysterious_Platinum_Nitride_2006_comp} arrived at an astonishing result: the experimentally proposed PtN(B3) is an entirely unstable structure at any pressure.

To closely study the non-zero pressure stoichiometric and structural preferences, we displayed in Fig. \ref{mixture_PtmNn_hp_EOS} the enthalpy $H$ \textit{vs.} pressure $P$ equation of states (EOS) for the most favorable Pt$_3$N phase (RhF$_3$), the three previously proposed PtN modifications (B1, B3 and B17), and the four considered PtN$_2$ structures in the present work (C1, C2, C18 and CoSb$_2$). The arrows indicate the pressures at which curves cross each other. From these curves, it is clear that PtN$_2$(C18), followed by PtN$_2$(C2), are the most energetically favorable phases at all pressure. At pressures above $10 \; \text{GPa}$, PtN$_2$(CoSb$_2$) has lower enthalpy than the rest of the modifications, including PtN(B17) and PtN$_2$(C1). At pressures higher than $8 \; \text{GPa}$, PtN(B17) becomes more favorable than Pt$_3$N(RhF$_3$), but the former never competes behind $52 \; \text{GPa}$ when PtN$_2$(C1) becomes more favorable. However, Pt$_3$N(RhF$_3$) is more stable than PtN(B3) at all pressures. In summary, Fig. \ref{mixture_PtmNn_hp_EOS} reveals that even if a PtN phase has been observed (at pressures around $50 \; \text{GPa}$), this phase must be unstable toward phase decomposition into solid constituents Pt and PtN$_2$ (see also Ref. \onlinecite{PtN_PtN2_2009_comp}) or into Pt and Pt$_3$N. However, the series of the possible phase transitions must be carefully investigated.
%
% -------------------------------------------------------------------------------------------------------------
\subsection{Volume per Atom and Lattice Parameters}\label{Volume per Atom and Lattice Parameters}
% -------------------------------------------------------------------------------------------------------------
The obtained equilibrium volume per atom $V_{0}$, i.e. the inverse of the number density, for all the considered modifications are numerally presented in Table \ref{platinum_nitrides_equilibrium_structural_properties} and graphically depicted relative to the Pt(A1) in Fig. \ref{platinum_nitrides_equilibrium_properties}. On average, Pt$_3$N phases tend not to change the number density of the host parent Pt(A1); PtN phases tend to slightly increase it; while the PtN$_2$ increase it significantly.

It is also evident from Fig. \ref{platinum_nitrides_equilibrium_properties} that in crossing the boarders between the Pt$_3$N and PtN and between the PtN and PtN$_2$ islands, i.e. in increasing the N content, $V_{0}$ tends to decrease while the volume per Pt atom $V_{0}^{\text{Pt}}$, a measure of the average Pt--Pt bond length, tends to increase. The latter finding has been found to be true for the nitrides of Cu \cite{Suleiman_PhD_arXiv2012_copper_nitrides_article} and Ag \cite{Suleiman_PhD_arXiv2012_silver_nitrides_article} as well.
%
% -------------------------------------------------------------------------------------------------------------
\subsection{\label{Bulk Modulus and its Pressure Derivative}Bulk Modulus and its Pressure Derivative}
% -------------------------------------------------------------------------------------------------------------
With only a few exceptions, Fig. \ref{platinum_nitrides_equilibrium_properties} and Table \ref{platinum_nitrides_equilibrium_structural_properties} reveal that nitridation of Pt apparently tends to reduce its bulk modulus. Relative to each other, the twenty $B_0$'s show no clear trend. The most energetically favorable PtN phase, B17, has $42 \; \text{GPa}$ higher bulk modulus than the proposed PtN(B3).

As we mentioned somewhere else \cite{Suleiman_PhD_arxive_silver_nitrides_article}, $B_0$ is far more sensitive to any change in volume than the change in $E_\text{coh}$. The case of PtN(B2) is a clear example, in which the slight decrease in $V_{0}$ overcomes the significant increase in $E_\text{coh}$ leading only to a very small decrease in $B_0$ (Fig. \ref{platinum_nitrides_equilibrium_properties} and Table \ref{platinum_nitrides_equilibrium_structural_properties}).

Given that all the considered phases have higher $E_\text{coh}$ than Pt(A1), the foregoing argument fails to explain the decrease in $B_0$ in the case of the structures which have lower $V_0$ than their parent Pt(A1) and have lower $E_\text{coh}$ than the extreme case PtN(B2). However, if one replaces $V_0$ in the argument above with $V_0^{\text{Pt}}$, the contradiction can be lifted. Therefore, we believe that the mechanical properties in these nitrides may be dominated by the effect of the Pt-Pt bond length more than the simple number density.

Although the GGA calculated $B_{0}$ values in the present and previous works (Table \ref{platinum_nitrides_equilibrium_structural_properties}) are far smaller than the reported experimental value, our obtained bulk modulus for PtN$_2$(C1) is $20 \; \text{GPa}$ higher than that of Pt(A1). This is exactly the measured value for Pt after the PtN formation took place. The observation was considered by Gregoryanz et al. as an indication that some N is dissolved in Pt \cite{first_PtN_2004_exp}. Recalling that the $B_0$ of the produced platinum nitride is $\sim 100 \; \text{GPa}$ than that of Pt(A1) \cite{first_PtN_2004_exp}, our GGA-obtained $B_0$ for PtN$_2$(C1) is $\sim 80 \; \text{GPa}$ less than the experimental value \footnote{Recall that we only consider values relative to Pt(A1) to eliminate systematic errors.}.

It may be worth to notice from Table \ref{platinum_nitrides_equilibrium_structural_properties} that the lattice parameter $a$ of PtN$_2$(C1) is $ 0.13 \; \text{\AA}$ higher than that of PtN(B3); yet the $B_0$ of the former is $\sim 70 \; \text{GPa}$ higher than the latter. This difference in $B_0$ can be attributed to the fact that in B3, N atoms occupy only half of the tetrahedral interstitial sites of the Pt sub-lattice, while in C1, the four remaining tetrahedral interstitial sites are filled with N atoms \cite{PtN2_2005_comp,Mysterious_Platinum_Nitride_2006_comp,PtN_PtN2_2005_comp}. This filling significantly reduces the compressibility but slightly increases the volume of the unit cell. This fact can also be seen readily as a consequence of the difference in the average volume per atom in the two cases (Table \ref{platinum_nitrides_equilibrium_structural_properties}).\\

The pressure derivative of the bulk modulus, $B_{0}^{\prime}$, measures the sensitivity of $B_{0}$ to any external pressure. The top subfigure in Fig. \ref{platinum_nitrides_equilibrium_properties} reveals that the bulk moduli of Pt$_3$N(Fe$_3$N) and PtN$_2$(C2, C18 and CoSb$_2$) increase upon application of external pressure. Pt$_3$N(D0$_2$) is very sensitive and its $B_{0}$ will increase significantly under an infinitesimal excess of pressure. Pt$_3$N(L1$_2$ and RhF$_3$) tend to be inert; while Pt$_3$N(D0$_3$, A15 and D0$_9$), PtN$_2$(C1) and all PtN phases tend to decrease their bulk modulus upon application of external pressure. Although $B_{0}^{\prime}$ is a measurable quantity \cite{B_prime_1997_theory_comp_n_exp}, we couldn't find any experimental value to test our obtained values against.
%
% -------------------------------------------------------------------------------------------------------------
\subsection{\label{Formation Energies}Formation Energies}
% -------------------------------------------------------------------------------------------------------------
From Fig. \ref{platinum_nitrides_equilibrium_properties} and Table \ref{platinum_nitrides_equilibrium_structural_properties}, it is evident that formation energy $E_f$ has the same trend as the cohesive energy $E_{\text{coh}}$. If $E_f$ is taken as a measure of synthesized, then the relatively most favorable Pt$_3$N phases have the same synthesized as the most favorable PtN and PtN$_2$.

A positive value of $E_f$ means, in principle, that, at the temperature and pressure at which $E_f$ is calculated, the phase is thermodynamically unstable (endothermic) and have a tendency to decompose into its constituent components. In our case, this observation is corroborated by the experimental fact that the synthesis of the platinum nitrides was achieved only at high temperature and temperature \cite{positive_Ef_n_PtN2_2006_comp_Scandolo,AuN2_and_PtN2_2010_comp}.

Using different methods, other researchers \cite{PtN_n_PtN2_2006_exp_n_comp,positive_Ef_n_PtN2_2006_comp_Scandolo,
AuN2_and_PtN2_2010_comp} also obtained positive (zero-pressure and zero-temperature) formation energies for some PtN and/or PtN$_2$ phases. Some of their values are included in Table \ref{platinum_nitrides_equilibrium_structural_properties} with indication to the methods of calculations.

The obtained relative difference in $E_f$ for PtN$_2$(C1) and PtN$_2$(C2) is in good agreement with Ref. \onlinecite{PtN_n_PtN2_2006_exp_n_comp}. However, the differences in our and their obtained $E_f$ values can be attributed to three factors: First, the difference in the obtained lattice parameter (see Table \ref{platinum_nitrides_equilibrium_structural_properties}). Second, the value of our calculated equilibrium free parameter $u$ is $0.417$ while Ref. \onlinecite{PtN_n_PtN2_2006_exp_n_comp} obtained $0.415$ \footnote{Fixing the lattice parameter at the experimental value $a = 4.8041 \; \text{\AA}$, Ref. \onlinecite{positive_Ef_n_PtN2_2006_comp_Scandolo} relaxed the N ions and obtained the same value $u = 0.415$.}. Third, and the most significant source of difference, the experimental value of $E_\text{coh}(\text{N}_2^{\text{gas}})$ in Eq. \ref{formation energy equation} was used by Ref. \onlinecite{PtN_n_PtN2_2006_exp_n_comp}, while we calculated it as described in sub-section \ref{Formation Energy}.

It may be worth mentioning here that a negative theoretical value of $E_f = - 0.4 \; eV/\text{atom}$ was obtained for PtN$_2$(C2) at $P = 50 \; \text{GPa}$, showing excellent agreement with experiment \cite{PtN_n_PtN2_2006_exp_n_comp}. Moreover, Young et al. \cite{positive_Ef_n_PtN2_2006_comp_Scandolo} claimed that PtN$_2$ dissociates upon mild heating below $P = 10 \; \text{GPa}$.
%
% -------------------------------------------------------------------------------------------------------------
\subsection{\label{Electronic Properties}Electronic Properties}
\begin{figure*}[H] % ----------------------------------------------------------------------------------
%\begin{center}
% Figure environments same as 0.8 * \textwidth please
\includegraphics[width=1.0\textwidth]{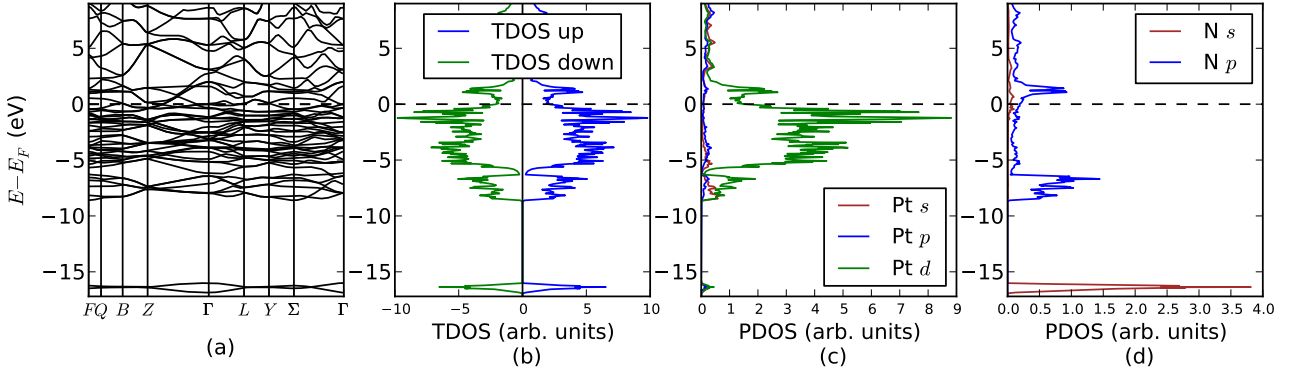}
\caption{\label{Pt3N1_RhF_3_electronic_structure}(Color online.) DFT calculated electronic structure for Pt$_3$N in the RhF$_3$ structure:
\textbf{(a)} band structure along the high-symmetry $\mathbf{k}$-points which are labeled according to Ref. [\onlinecite{Bradley}]. Their coordinates w.r.t. the reciprocal lattice basis vectors are: $F(0.5, 0.5, 0.0)$, $Q(0.375, 0.625, 0.0)$, $B(0.5, 0.75, 0.25)$, $Z(0.5, 0.5, 0.5)$, $\Gamma(0.0,  0.0, 0.0)$, $L(0.0, 0.5, 0.0)$, $Y(0.25, 0.5, -.25)$, $\Sigma(0.0, 0.5, -.5)$;
 \textbf{(b)} spin-projected total density of states (TDOS);
 \textbf{(c)} partial density of states (PDOS) of Pt($s, p, d$) orbitals in Pt$_3$N;
 and \textbf{(d)} PDOS of N($s, p$) orbitals in Pt$_3$N.}
%\end{center}
\end{figure*}

\begin{figure*}[H]	% ------------------------------------------------------------------------------------
%\begin{center}
% Figure environments same as 0.8 * \textwidth please
\includegraphics[width=1.0\textwidth]{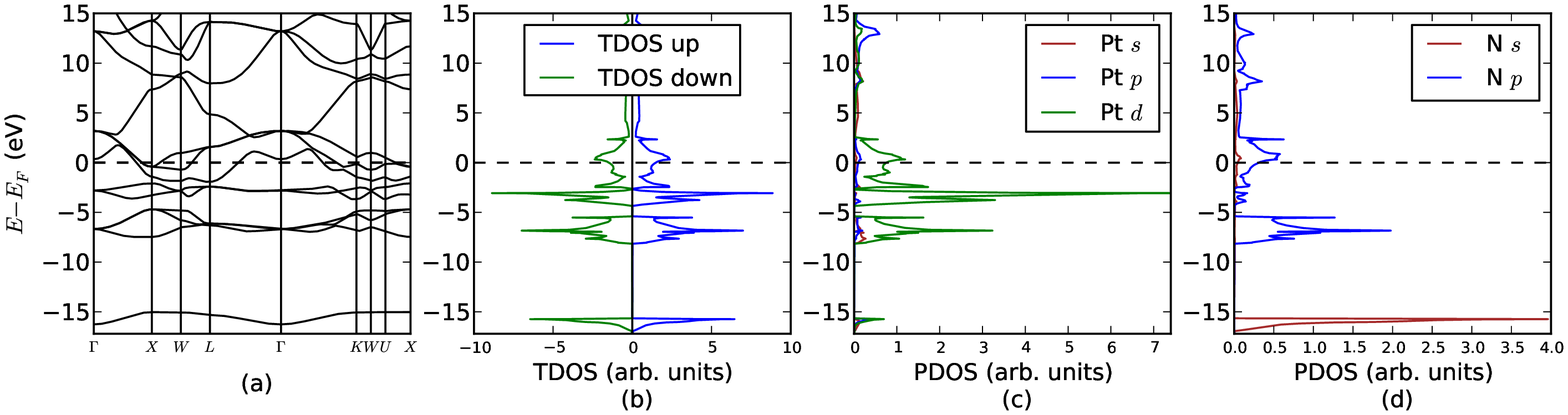}
\caption{\label{Pt1N1_B3_electronic_structure}(Color online.) DFT calculated electronic structure for PtN in the B3 structure:
 \textbf{(a)} band structure along the high-symmetry $\mathbf{k}$-points which are labeled according to Ref. [\onlinecite{Bradley}] \footnote{\textit{\textbf{The coordinates of the W point is not as the same as in Ref. [\onlinecite{Bradley}], but they are equivalent. Also, the coordinates of U and K are not given in Ref. [\onlinecite{Bradley}]. The coordinates of U, K and the equivalent W were created by means of \href{http://www.xcrysden.org/}{XCrySDen}!}}}. Their coordinates w.r.t. the reciprocal lattice basis vectors are: $\Gamma (0.0, 0.0, 0.0)$, $X (0.0, 0.5, 0.0)$, $W (0.75, 0.25, 0.5)$, $L (0.5, 0.5, 0.5)$, $K (0.750, 0.375, 0.375)$, $U (0.625, 0.250, 0.625)$;
 \textbf{(b)} spin-projected total density of states (TDOS);
 \textbf{(c)} partial density of states (PDOS) of Pt($s, p, d$) orbitals in PtN;
 and \textbf{(d)} PDOS of N($s, p$) orbitals in PtN.
}
%\end{center}
\end{figure*}

\begin{figure*}[H]	% ------------------------------------------------------------------------------------
%\begin{center}
% Figure environments same as 0.8 * \textwidth please
\includegraphics[width=1.0\textwidth]{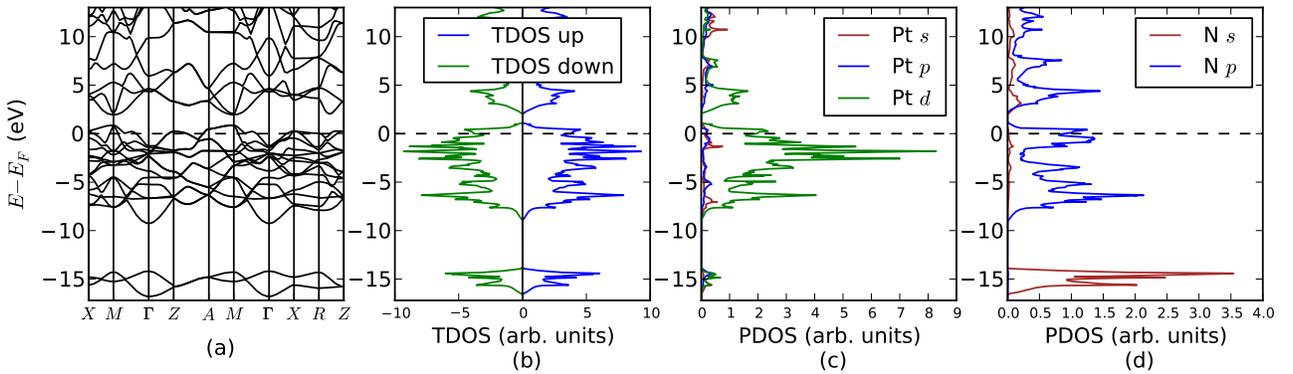}
\caption{\label{Pt1N1_B17_electronic_structure}(Color online.) DFT calculated electronic structure for PtN in the B17 structure:
 \textbf{(a)} band structure along the high-symmetry $\mathbf{k}$-points which are labeled according to Ref. [\onlinecite{Bradley}]. Their coordinates w.r.t. the reciprocal lattice basis vectors are: $X (0.0, 0.5, 0.0)$, $M (0.5, 0.5, 0.0)$, $\Gamma (0.0, 0.0, 0.0)$, $Z (0.0, 0.0, 0.5)$, $A (0.5, 0.5, 0.5)$, $R (0.0, 0.5, 0.5)$;
 \textbf{(b)} spin-projected total density of states (TDOS);
 \textbf{(c)} partial density of states (PDOS) of Pt($s, p, d$) orbitals in PtN;
 and \textbf{(d)} PDOS of N($s, p$) orbitals in PtN.}
%\end{center}
\end{figure*}

\begin{figure*}[H]	% -------------------------------------------------------------------------
%\begin{center}
% Figure environments same as 0.8 * \textwidth please
\includegraphics[width=1.0\textwidth]{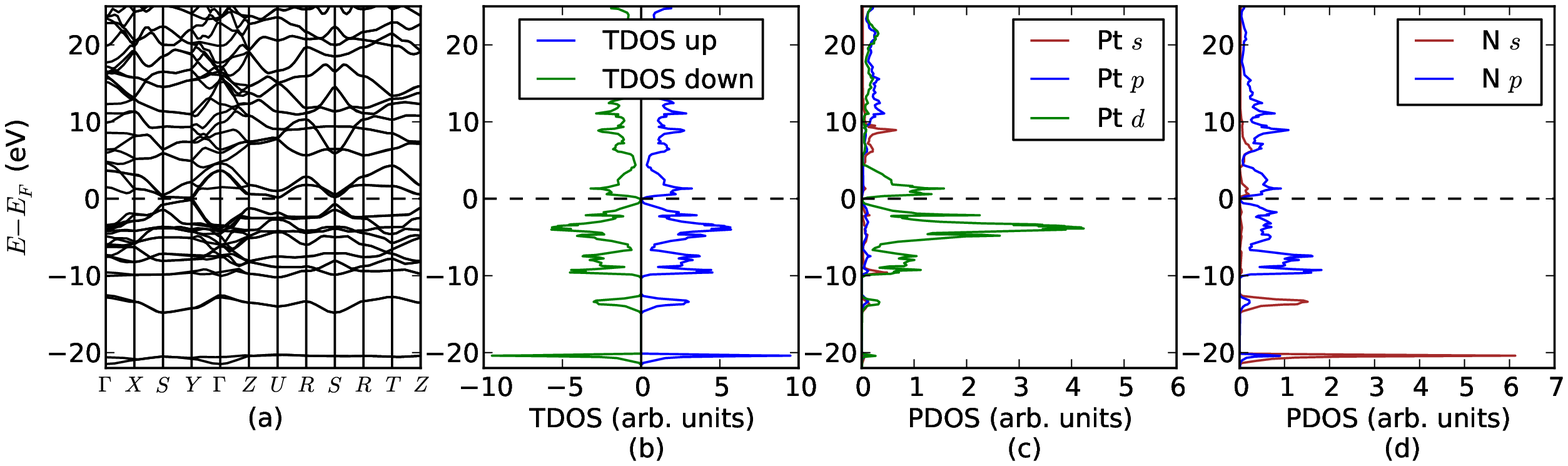}
\caption{\label{Pt1N2_C18_electronic_structure}(Color online.) DFT calculated electronic structure for PtN$_2$ in the C18 structure:
 \textbf{(a)} band structure along the high-symmetry $\mathbf{k}$-points which are labeled according to Ref. [\onlinecite{Bradley}]. Their coordinates w.r.t. the reciprocal lattice basis vectors are: $\Gamma( 0.0, 0.0, 0.0)$, $X( 0.0, 0.5, 0.0)$, $S( -.5, 0.5, 0.0)$, $Y( -.5, 0.0, 0.0)$, $Z( 0.0, 0.0, 0.5)$, $U( 0.0, 0.5, 0.5)$, $R( -.5, 0.5, 0.5)$, $T( -.5, 0.0, 0.5)$;
 \textbf{(b)} spin-projected total density of states (TDOS);
 \textbf{(c)} partial density of states (PDOS) of Pt($s, p, d$) orbitals in PtN$_2$;
 and \textbf{(d)} PDOS of N($s, p$) orbitals in PtN$_2$.}
%\end{center}
\end{figure*}

% -------------------------------------------------------------------------------------------------------------
The DFT obtained band diagrams $\epsilon_{i}^{\sigma}(\mathbf{k})$ and spin-projected total and partial density of states (DOS) of the most stable modifications: Pt$_3$N(RhF$_3$), PtN(B3 and B17), and PtN$_2$(C18) are displayed in Figs. \ref{Pt3N1_RhF_3_electronic_structure}, \ref{Pt1N1_B3_electronic_structure}, \ref{Pt1N1_B17_electronic_structure} and \ref{Pt1N2_C18_electronic_structure}, respectively. Spin-projected total density of states (TDOS) are shown in sub-figure (b) in each case. Because in these four considered cases electrons occupy the spin-up and the spin-down bands equally, it was sufficient only to display spin-up DOS and spin-up band diagrams. Displaying the energy bands along densely sampled high-symmetry strings of $\mathbf{k}$-points allows us to extract information about the electronic structure of these phases. Moreover, to investigate the details of the orbital character of the bands, the Pt($s, p, d$) and N($s, p$) resolved DOS's are plotted at the same energy scale.
% -----------------------------------------------------------------------------------------------------
%
	% ------------------------------------------------------------------------------------------
	% ------------------------------------------------------------------------------------------
	% -------------------------------------------------------------------------------

With The Fermi surface crossing the partly occupied bands. it is clear from Figs. \ref{Pt3N1_RhF_3_electronic_structure}, \ref{Pt1N1_B3_electronic_structure}, and \ref{Pt1N1_B17_electronic_structure} that Pt$_3$N(RhF$_3$) PtN(B3) and PtN(B17) are metals.

The TDOS of Fig. \ref{Pt1N2_C18_electronic_structure}(b) reveal that PtN$_2$(C18) is a semiconductor with (Fig. \ref{Pt1N2_C18_electronic_structure}(a)) its valence band maximum (VBM) at $(Y,-0.091 \; eV)$ and its conduction band minimum (CBM) at $(Y,0.044 \;eV)$, resulting in a narrow direct energy band gap $E_g = 0.135 \; eV$ of width. Below this fundamental gap there are three bands: the deep one at $\sim -20.5 \; eV$ consists mainly of the N$(2s)$ states. Its high DOS and sharp feature correspond to its little and slow energy variation in the $\mathbf{k}-$space. The second band is relatively narrow ($\sim 2.2 \; eV$ of width) with low density and steming mainly from a mixture of the N states with Pt($d$) states. The superposition Pt($d$) and N($p$) states in the region from $-10.314 \; eV$ to $-0.091 \; eV$ below the fundamental gap constitutes the third band with highly structured, intense and narrow series of peaks. Our obtained TDOS and PDOS show excellent agreement with Ref. \onlinecite{AuN2_and_PtN2_2010_comp} where also PtN$_2$(C18) was predicted to be a semiconductor, but band diagrams and $E_g$ value are not given.

It may be worth mentioning here that PtN(B1) \cite{PtN_2006_comp} and PtN(B4) \cite{PtN_2007_comp} were found to be metallic, PtN$_2$(C1) was found to be a poor metal \cite{PtN_PtN2_2005_comp}, PtN$_2$(CoSb$_2$) \cite{AuN2_and_PtN2_2010_comp} was found to be a semiconductor, and an indirect band gap between $1.2 \; eV$ \cite{AuN2_and_PtN2_2010_comp} and $1.5 \; eV$ \cite{positive_Ef_n_PtN2_2006_comp_Scandolo} has been obtained for PtN$_2$(C2). 
%
% -------------------------------------------------------------------------------------------------------------
\subsection{\label{Optical Properties}Optical Properties}
\begin{figure*}	% --------------------------------------------------------------------------------
%\begin{center}
% Figure environments same as 0.8 * \textwidth please
%height=0.10\textheight
\includegraphics[width=1.0\textwidth]{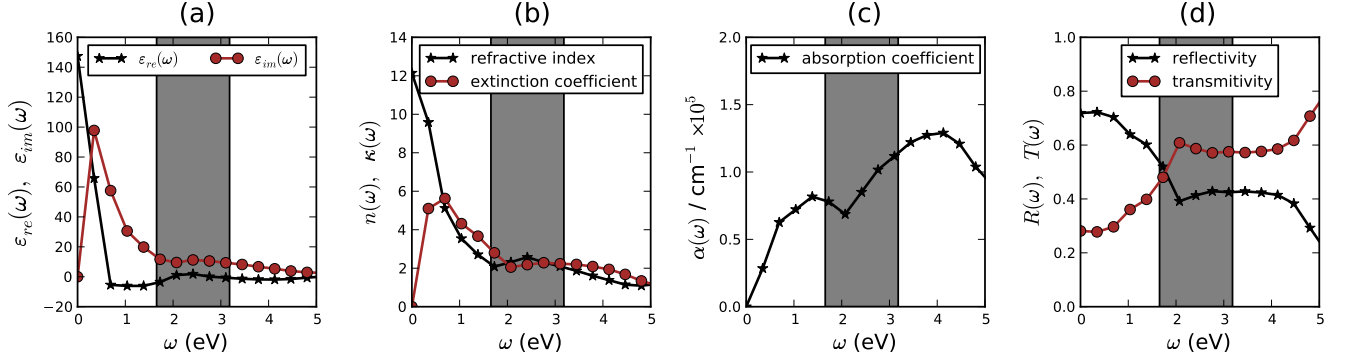}
\caption{\label{Pt1N1_B3_optical_constants}(Color online.) 
The $GW$ calculated frequency-dependent optical spectra of PtN(B3):
 \textbf{(a)} the real $\varepsilon_{\text{re}}(\omega)$ and the imaginary $\varepsilon_{\text{im}}(\omega)$ parts of the dielectric function $\varepsilon_{\text{RPA}}(\omega)$;
 \textbf{(b)} refraction $n(\omega)$ and extinction $\kappa(\omega)$ coefficients;
 \textbf{(c)} absorption coefficient $\alpha(\omega)$; and 
 \textbf{(d)} reflectivity $R(\omega)$ and transmitivity $T(\omega)$.
 The shaded window highlights the optical region.}
%\end{center}
\end{figure*}

\begin{figure*}	% --------------------------------------------------------------------------------
\includegraphics[width=1.0\textwidth]{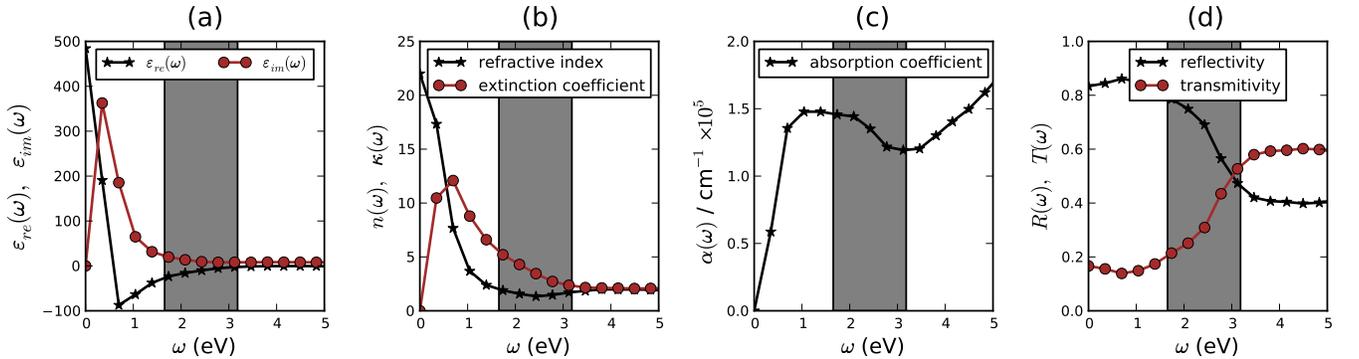}
\caption{\label{Pt1N1_B17_optical_constants}(Color online.) 
The $GW$ calculated frequency-dependent optical spectra of PtN(B17):
 \textbf{(a)} the real $\varepsilon_{\text{re}}(\omega)$ and the imaginary $\varepsilon_{\text{im}}(\omega)$ parts of the dielectric function $\varepsilon_{\text{RPA}}(\omega)$;
 \textbf{(b)} refraction $n(\omega)$ and extinction $\kappa(\omega)$ coefficients;
 \textbf{(c)} absorption coefficient $\alpha(\omega)$; and 
 \textbf{(d)} reflectivity $R(\omega)$ and transmitivity $T(\omega)$.
 The shaded window highlights the optical region.}
\end{figure*}

% -------------------------------------------------------------------------------------------------------------
$GW$ calculations were carried out for the PtN(B3) and PtN(B17) metallic phases at their equilibrium. Figs. \ref{Pt1N1_B3_optical_constants} and \ref{Pt1N1_B17_optical_constants} display the obtained real and imaginary parts of the frequency-dependent dielectric function $\varepsilon_{\text{RPA}}(\omega)$ of these two phases and the corresponding derived optical spectra (Eqs. \ref{n(omega)} -- \ref{R(omega)}). In each sub-figure, the optical region $\left[\sim (3.183 - 1.655) \; eV  \equiv  (390 - 750) \; nm\right]$ is shaded.
	% ------------------------------------------------------------------------------------
	% ------------------------------------------------------------------------------------

The non-vanishing absorption coefficient $\alpha\left(\omega\right)$ in the whole range for both phases confirms their metallic character. As it should be the case, refraction $n(\omega)$ and extinction $\kappa(\omega)$ coefficients behave as the real $\varepsilon_{\text{re}}(\omega)$ and the imaginary $\varepsilon_{\text{im}}(\omega)$ dielectric functions, respectively.

As one can see from sub-figure \ref{Pt1N1_B3_optical_constants}(d), close to the edge of the optical region at $\sim (1.762 \; eV  =  703.768 \; nm)$ PtN(B3) is $50 \%$ reflector and $50 \%$ transmitter. From $\sim (2.071 \; eV  =  598.579 \; nm)$ to the UV region, PtN(B3) is only $\sim 40 \%$ reflecting but $\sim 60 \%$ transmitting. However, more of the transmitted portion in this region will be absorbed as the photon energy increases. This fact can be readily noticed if one compares sub-figures \ref{Pt1N1_B3_optical_constants}(c) and \ref{Pt1N1_B3_optical_constants}(d).

PtN(B17), as can be seen from sub-figure \ref{Pt1N1_B17_optical_constants}(d), is a very good reflector in the whole region until $\sim (3.000 \; eV = 413.281 \; nm)$ where it equally reflects and transmits the violet light. However, less of the transmitted portion in the optical region will be absorbed as the photon wavelength decreases. This fact can be readily observed in sub-figures \ref{Pt1N1_B17_optical_constants}(c).\\

According to the best of our knowledge, there is no available experimental optical spectra for the platinum nitride. However, from their visual appearance, all the synthesized platinum nitride samples look very shiny and darker than their parent platinum in reflected light and totally opaque in transmitted light. These features suggest that PtN is either a poor metal or a semiconductor with a small band gap \cite{first_PtN_2004_exp}.

From Figs. \ref{Pt1N1_B17_optical_constants} and \ref{Pt1N1_B17_electronic_structure}, the above mentioned properties are strongly met by PtN(B17), but purely seen (Figs. \ref{Pt1N1_B3_optical_constants} and \ref{Pt1N1_B3_electronic_structure}) in PtN(B3), as discussed above. Unfortunately, we did not carry out optical calculations for PtN$_2$(C1 or C2).
%
% -------------------------------------------------------------------------------------------------------------
\subsection{\label{PtN versus PtN$_2$}PtN versus PtN$_2$}
% -------------------------------------------------------------------------------------------------------------
Using our own obtained results in the present work as well as the findings of other researchers, below we make a comparison between the PtN modifications (supported by the experimentalists) and the PtN$_2$ phases (supported by the theoreticians):

\begin{itemize}	% ------------------------------------------------------------------------
\item Given that GGA calculated lattice parameters are usually overestimated \cite{accurate_GGA_2006,GGA_vs_LDA_2004,Ch1_in_Primer_in_DFT_2003}, the obtained values of the $a$ lattice parameter for PtN$_2$(C1 and C2) are the closest ones to the experimental value (to within $3\; \%$ and $2\; \%$, respectively), while the PtN phases are in poor agreement with experiment, as can be seen in Table \ref{platinum_nitrides_equilibrium_structural_properties}.

\item First-principles studies of transition metals nitrides show that the $B_0$'s of the elemental metals are  generally enhanced by nitridation \cite{first_PtN_2004_exp,trends+heavyTMNDFT1999}. Compared to experiment, Table \ref{platinum_nitrides_equilibrium_structural_properties} and Fig. \ref{platinum_nitrides_equilibrium_properties} reveal that this trend is met by PtN$_2$(C1), while PtN(B3) has $50 \; \text{GPa}$ lower than Pt(A1).

\item Like the first synthesized sample and the proposed PtN(B3) modification \cite{first_PtN_2004_exp}, PtN$_2$(C1 \cite{PtN_PtN2_2005_comp} and C2 \cite{PtN_n_PtN2_2006_exp_n_comp}) have fcc sub-lattice of Pt.

\item PtN$_2$(
C1 \cite{PtN_PtN2_2005_comp},
 C2 \cite{positive_Ef_n_PtN2_2006_comp_Scandolo,AuN2_and_PtN2_2010_comp}, 
 C18 \cite{AuN2_and_PtN2_2010_comp} and
 CoSb$_2$ \cite{AuN2_and_PtN2_2010_comp}) have all been found to be elastically stable, while PtN(B3\cite{PtN_n_PtN2_2005_July_comp_5+,PtN_PtN2_2005_comp,PtN_2006_comp,
positive_Ef_n_PtN2_2006_comp_Scandolo} and B17 \cite{PtN_2006_comp}) were found to be elastically unstable.

\item Formation and cohesive energies of PtN$_2$(C2, C18 and CoSb$_2$) are lower than that of PtN(B3) [Table \ref{platinum_nitrides_equilibrium_structural_properties} and Fig. \ref{platinum_nitrides_equilibrium_properties}].

\item In excellent agreement with experiment, the calculated formation energy of PtN$_2$(C2) at $P = 50 \; \text{GPa}$ was calculated to be negative \cite{PtN_n_PtN2_2006_exp_n_comp}, while calculations found PtN(B3) to be thermodynamically unstable at all pressures \cite{Mysterious_Platinum_Nitride_2006_comp}.

\item The experimentally obtained Raman spectrum of the reproduced platinum nitride \cite{PtN_n_PtN2_2006_exp_n_comp} matches closely that of pyrite (FeS2), i.e. in the C2 structure, but does not match the PtN(B3) spectrum that expected from group theory \cite{PtN_n_PtN2_2006_exp_n_comp}. 

\item The theoretically calculated \cite{positive_Ef_n_PtN2_2006_comp_Scandolo,AuN2_and_PtN2_2010_comp} Raman spectrum for PtN$_2$(C2) shows good agreement with the first experimentally obtained one \cite{first_PtN_2004_exp}. 

\item In agreement with the experimental observation and the visual appearance of the first produced platinum nitrides \cite{first_PtN_2004_exp}, PtN$_2$(C1) was found to be a poor metal \cite{PtN_PtN2_2005_comp}, and we found PtN$_2$(C18) to be a semiconductor with a small band gap.
\end{itemize} 	% ------------------------------------------------------------------------

Hence, in contrast to the proposed PtN modifications, PtN$_2$ phases possess many similar properties as the synthesized phase \footnote{Such an observation was arrived at by other authors \cite{PtN_PtN2_2005_comp} for the PtN(C1). Here we are making more comprehensive comparison}.
%
% =============================================================================================================
\section{\label{Conclusions}Conclusions}
% =============================================================================================================
In summary, we presented a systematic series of first-principles calculations of the energy-optimized geometries, phase stabilities and electronic and optical properties of bulk Pt$_3$N, PtN and PtN$_2$ in twenty different crystal structures. Comprehensive comparison with experiment and with previous calculations has been made, and excellent agreement has been achieved. We found that both the lowest energy and the highest bulk modulus phases belong to the PtN$_2$ series and not to the PtN family. Moreover, the calculated electronic and optical properties of the PtN$_2$ phases show stronger consistency with experiment than the claimed PtN(B3) phase. In the present work, we have investigated a wider parameter sub-space than previous calculations, and to the best of our knowledge, the present work is the first to propose and to study the physical properties of Pt$_3$N, as well as the first to theoretically calculate the optical spectra of this new material. However, optical properties of PtN$_2$(C1 and C2) have not been investigated, and we strongly recommend optical calculations for these phases and obtained results should be tested against experiment. Moreover, experimentalists should provide the community with more data.
%
% =============================================================================================================
\section*{Acknowledgments}
% =============================================================================================================
We thank the CHPC for providing the supercomputer facilities to perform the GW calculations. Suleiman would like to acknowledge the support he received from Wits, DAAD, AIMS, SUST, and the ASESMA and the KWAMS13 groups.
%\nocite{*}

\bibliography{v1_arXiv_platinum_nitrides_article}
\end{document}